\documentclass[12pt]{article}

\usepackage[dvips]{graphics}  
\usepackage{amssymb}

\newcommand{\nc}{\newcommand}  
\nc{\figcap}[1]{\begin{quote}\refstepcounter{figure}  
        {\bf Figure \thefigure}: {\small #1}\end{quote}}  
\nc{\fig}[1]{\mbox{Fig.~\ref{#1}}}

\nc{\noi}{\noindent}

\nc{\bea}{\begin{eqnarray}}  
\nc{\eea}{\end{eqnarray}}  
\nc{\bean}{\begin{eqnarray*}}  
\nc{\eean}{\end{eqnarray*}}  
\nc{\ba}{\begin{array}}  
\nc{\ea}{\end{array}}  
\nc{\be}{\begin{equation}}  
\nc{\ee}{\end{equation}}  
\nc{\nn}{\nonumber}  
\nc{\bra}[1]{\langle #1|}  
\nc{\ket}[1]{|#1\rangle}  
\nc{\av}[1] {\langle #1\rangle}  
\nc{\vac}[1] {\langle 0| #1|0\rangle}  
\nc{\amp}[2]{\langle #1|#2\rangle}  
\nc{\da}{\dagger}  
\nc{\pa}{\partial}  
\nc{\ga}{\gamma}  
\nc{\ep}{\epsilon}  
\nc{\tf}{t_f}  
\nc{\half}{\ensuremath{\frac{1}{2}}}  
\nc{\hHH}{\hat H}  
\nc{\ha}{\hat a}  
\nc{\hO}{\hat O}  
\nc{\hAA}{\hat A}  
\nc{\hB}{\hat B}  
\nc{\hG}{\hat G}  
\nc{\hN}{\hat N}  
\nc{\hU}{\hat U}  
\nc{\hx}{\hat{x}}  
\nc{\hp}{\hat{p}}  
\nc{\hpsi}{\hat \psi}  
\nc{\hphi}{\hat \phi}  
\nc{\hpi}{\hat \pi}  
\nc{\hpd}{\hat \psi ^\dagger}  
\nc{\hE}{\hat E}  
\nc{\hb}{\hat b}  
\nc{\hc}{\hat c}  
\nc{\hjo}{\hat j _0}  
\nc{\hrho}{\hat \rho}  
\nc{\leave}{\! \! \! \! \! / \, \,}  
\nc{\intl}[1]{\int d\! #1 \,} 
\nc{\intll}[3]{\int _#1^#2 d\! #3 \,} 
\nc{\lm}{\lim _{y \rightarrow x}}  
  
\nc{\scd}{\partial ^2 _{A_T}}  
\nc{\fd}[1]{\frac{\delta }{\delta #1}} 
\nc{\pad}[1]{\frac{\partial}{\partial #1}} 
\nc{\refpa}[1]{(\ref{#1})} 

\nc{\calH}{\ensuremath{\mathcal{H}}}  
\nc{\calD}{\ensuremath{\mathcal{D}}}  
\nc{\calL}{\ensuremath{\mathcal{L}}}  
\nc{\calO}{\ensuremath{\mathcal{O}}}  
\nc{\hcalO}{\ensuremath{\hat \mathcal{O}}}  
\nc{\calK}{\ensuremath{\mathcal{K}}}  
\nc{\Tr}{\ensuremath{\mathrm{Tr}}}  
\nc{\tr}{\ensuremath{\mathrm{tr}}}  
\nc{\ra}{\rightarrow}  
\nc{\lr}{\leftrightarrow}  
\nc{\phistar}{\phi^*}  
\nc{\etat}{\eta_T}  
\nc{\het}{\hat E_T}  
\nc{\hpt}{\hat \psi_T}  
\nc{\hpdt}{\hat \psi ^\dagger_T}  
\nc{\bart}{\bar{t}}  
\nc{\barp}{\bar{p}}  
\nc{\barT}{\bar{T}}  
\nc{\hbarrho}{\hat{\bar{\rho}}}  
\nc{\bga}{\ensuremath{\mbox{\boldmath{$\gamma$}}}}  
\nc{\bsi}{\ensuremath{\mathbf{\sigma}}}  
\nc{\bx}{\ensuremath{\mathbf{x}}}  
\nc{\by}{\ensuremath{\mathbf{y}}}  
\nc{\bz}{\ensuremath{\mathbf{z}}}  
\nc{\bp}{\ensuremath{\mathbf{p}}}  
\nc{\bn}{\ensuremath{\mathbf{n}}}  
\nc{\bbp}{\ensuremath{\bar{\mathbf{p}}}}  
\nc{\bP}{\ensuremath{\mathbf{P}}}  
\nc{\hbA}{\hat{\ensuremath{\mathbf{A}}}}  
\nc{\hbB}{\hat{\ensuremath{\mathbf{B}}}}  
\nc{\bA}{\ensuremath{\mathbf{A}}}  
\nc{\bJ}{\ensuremath{\mathbf{J}}}  
\nc{\bB}{\ensuremath{\mathbf{B}}}  
\nc{\bH}{\ensuremath{\mathbf{H}}}  
\nc{\bM}{\ensuremath{\mathbf{M}}}  
\nc{\bD}{\ensuremath{\mathbf{D}}}  
\nc{\bE}{\ensuremath{\mathbf{E}}}  
\nc{\hbE}{\hat{\ensuremath{\mathbf{E}}}}  
\nc{\br}{\ensuremath{\mathbf{r}}}  
\nc{\bj}{\ensuremath{\mathbf{j}}}  
\nc{\bOm}{\ensuremath{\mathbf{\Om}}}  
\nc{\om}{\omega}  
\nc{\Om}{\Omega}  
\nc{\sgn}{\mbox{sgn}}  
\nc{\deltabar}{\mbox{$\delta\hspace*{-8pt}\vspace*{-8pt}-$}}  
\nc{\gammat}{\tilde{\gamma}}  
\nc{\binom}[2] {{#1\choose #2}}  
\nc{\mub}{\bar{\mu}}  
\nc{\rhob}{\bar{\rho}}  
\nc{\Bb}{\bar{B}}  
\nc{\Jb}{\bar{J}}  
\nc{\Mb}{\bar{M}}  
\nc{\Tb}{\bar{T}}  
\nc{\sbar}{\bar{s}}  
\nc{\betab}{\bar{\beta}}  
\nc{\hj}{\hat j}  
\nc{\hQ}{\hat Q}  
\nc{\hJ}{\hat J}  
\nc{\hA}{\hat A}  
\nc{\hH}{\hat H}  
\nc{\de}{\delta}  
\nc{\leri}{\leftrightarrow}  
\nc{\llabel}[1]{\label{#1}\marginpar{#1}}

\nc{\bc}{\begin{center}}  
\nc{\ec}{\end{center}}  
\nc{\inv}[1]{\frac{1}{#1}}

\newlength{\overeqskip}  
\newlength{\undereqskip}  
\nc{\eq}[1]{\mbox{Eq.~(\ref{#1})}}  
\nc{\eps}{\epsilon}  
\nc{\goto}{\rightarrow}

\nc{\cF}{{\cal F}}  
\nc{\cG}{{\cal G}}  
\nc{\cH}{{\cal H}}

\begin{document}  
%
%
%
%
%
%
\thispagestyle{empty}   
%
%
%
%
{\begin{tabular}{lr}   
%
%
%
\end{tabular}}   
%
%
\vspace{10mm}   
\begin{center}   
\baselineskip 1.2cm   
{\Huge\bf   Noise and Order in Cavity Quantum Electrodynamics  
}\\[1mm]   
\normalsize   
\end{center}   
\vspace{1cm}  
{\centering   
{\large Per K. Rekdal\footnote{Email address: perr@phys.ntnu.no.}$^{}$   
and   
{\large Bo-Sture K. Skagerstam\footnote{Email   
address: boskag@phys.ntnu.no.}$^{}$}   
 \\[5mm]   
  
$^{}~$Department of Physics, The Norwegian University of Science and Technology,     
N-7491 Trondheim, Norway \\[1mm]  
} }   
%
%
%
\vspace{1cm}


\begin{abstract}   
\normalsize   
%
%
\vspace{5mm}

   In this paper we investigate the various aspects of noise and order in the micromaser system.  
   In particular, we study the effect of adding fluctuations to the atom cavity transit time or   
   to the atom-photon frequency detuning. By including such noise-producing mechanisms we study the   
   probability and the joint probability for excited atoms to leave the cavity.   
   The influence of such fluctuations on the phase structure of the micromaser as well as on the   
   long-time atom correlation length is also discussed. 
   We also derive the asymptotic form of micromaser observables.

\end{abstract}


\newpage

\vspace{0cm}  
\bc{  
\section{Introduction}  
}\ec  
\vspace{0.5cm}

    Noise is usually considered as a limiting factor in the performance of a physical device  
    (see e.g. Refs.\cite{noises}).   
    There are, however, nonlinear dynamical systems where the presence of noise sources   
    can induce completely new regimes that cannot be realized without noise.  
    Recent studies have shown that noise in such systems can induce more   
    ordered regimes, more regular structures, increase the  
    degree of coherence, cause the amplification of weak signals and growth of their   
    signal-to-noise ratio (see e.g. Refs. \cite{Maritan&B&K94,Buchleitner98}).  
    In other words, noise can play a constructive role, enhancing  
    the degree of order in a system.

    The micromaser is an example of such a nonlinear system.   
    The micromaser system is an experimental realization of the idealized  
    system of a two-level atom interacting with a second quantized single-mode of the  
    electromagnetic field (for reviews and references see e.g. Refs. \cite{Walther_div}- \cite{Skagerstam99}).  
    In the micromaser a beam of two-level atoms is sent through a microcavity where each atom intersects   
    with the photon field inside the cavity during a transit time $\tau$.  
    After exit from the cavity the atoms are detected in either of its two states.  
    It is assumed that subsequent atoms arrive at time intervals which are much longer then   
    the atom-field interaction such that at most one atom at a time is inside the cavity,  
    which is the operating condition for the one-atom maser. In such a system noise-controlled  
    jumps between metastable states have been discussed in the literature \cite{Buchleitner98}.

    In the present paper we study the effect of including noise-producing mechanisms   
    in the micromaser system  
    like a velocity spread in the atomic beam or a spread in the atom-photon frequency detuning  
    $\Delta \omega$. We show that under suitable 
    conditions, such noise-producing mechanisms lead to more pronounced revivals in the   
    probabilities ${\cal P}(+)$ and   
    ${\cal P}(+,+)$, where ${\cal P}(+)$ is the probability that an atom is found in its excited   
    state after interaction with the photon field inside the microcavity and ${\cal P}(+,+)$ is the   
    joint probability that two consecutive atoms are measured in their  
    excited state after interaction.   
    It is the purpose of the present paper to study the counterintuitive role that noise   
    can play in the micromaser and extend the results in Refs. \cite{Filipowicz86_igjen,ElmforsLS95}.

   The paper is organized as follows.   
   For the convenience of the reader we recapitulate in Section \ref{revivals_SEC}    
   the theoretical framework of the micromaser system. In this Section we also discuss the   
   general conditions to make revivals in the probability ${\cal P}(+)$ well separated in   
   the atomic transit time.   
   In Section \ref{no_avg_SEC} we determine asymptotic limits of the order parameter   
   of the micromaser system and the probabilities ${\cal P}(+)$ and ${\cal P}(+,+)$.   
   In Section \ref{fluctuations_SEC} the effect of noise on ${\cal P}(+)$ and ${\cal P}(+,+)$,  
   the order parameter and the micromaser phase diagram is discussed.  
   In Section \ref{corr_KAP} the effect of fluctuations on the correlation length is discussed and  
   numerical investigations are presented.   
   A conclusion is given in Section \ref{final_KAP}.

\vspace{0.8cm}  
  
\bc{  
\section{The Dynamical System}  
\label{revivals_SEC} }  
\ec  
  
\vspace{0cm}

    In the description of the dynamics of the one-atom micromaser we have to take    
    losses of the photon field into account,  
    i.e. the time evolution of the photon field is described by a master equation.  
    The continuous time formulation of the micromaser system \cite{Lugiato87} is a suitable   
    technical frame work for our purposes.  
    Let $a$ be the probability for a pump atom to be in its excited state. Assuming    
    that the pump atoms are prepared in an incoherent  
    mixture, i.e. the density matrix of the atoms is diagonal with diagonal matrix elements   
    $a$ and $b$ such that $a+b=1$, it is shown in \cite{Lugiato87,ElmforsLS95} that   
    the vector $p$ formed by the diagonal density matrix elements of the photon  
    field obeys the differential equation $dp/dt=-\gamma Lp$. Here  
    $\gamma$ is the damping rate of photons in the cavity and  
    $L=L_C -N(M-1)$, where $( L_C)_{nm} = (n_b+1)[ \, n \delta_{n,m} - (n+1) \delta_{n+1,m} \, ] +    
    n_b[ \, (n+1)\delta_{n,m} - n \delta_{n,m+1} \, ]$ describes the damping of the cavity.  
    $n_b$ is the number of thermal photons and  
    $N$ is the average number of atoms injected into the cavity during the cavity decay time.  
    The matrix $M$ is $M=M(+) +M(-)$, where $M(+)_{nm} = b q_{n+1} \delta_{n+1,m} +   
    a(1-q_{n+1}) \delta_{n,m}$ and $M(-)_{nm} = a q_{n} \delta_{n,m+1} + b(1-q_{n})   
    \delta_{n,m}$, has its origin in the Jaynes-Cummings (JC) model \cite{Jaynes&Cummings63,ElmforsLS95}.  
    The quantity $q_n \equiv q(x=n/N)$ is given by

\be \label{q_n}  
    q(x) = \frac{x}{x+\Delta^2} \sin^2 \left ( \theta \sqrt{x+\Delta^2}\right )  ~~,  
\ee

    \noi  
    where $\Delta^2 \equiv (\Delta \omega)^2/(4g^2 N)$ is the scaled dimensionless detuning parameters of the   
    micromaser and  $g$ is the single photon Rabi frequency at zero detuning of the JC model.  
    \eq{q_n} is expressed in terms of the scaled dimensionless pump parameter $\theta = g \tau \sqrt{N}$.  
    Noise mechanisms, like a spread in the transition time or a spread in the   
    detuning, are now included in the analysis by simply averaging the matrix $L$ with respect   
    to $\theta$ or $\Delta$, i.e. averaging the quantity $q_n$ with respect to one of these parameters.   
    A similar averaging procedure with regard to the parameters $a$ (or $b$) or $n_b$ leads only to a trivial  
    replacement of the corresponding  parameters with their mean values.  
    The stationary solution of the photon distribution where such noise effects are included is  
    therefore derived in a standard and well known manner \cite{Filipowicz86_igjen,Lugiato87,ElmforsLS95}.  
    The result is

\be \label{p_n_eksakt}  
    {\bar p}_n = {\bar p}_0 \prod_{m=1}^{n}  \frac{n_b \, m +  N a \langle q_m \rangle}  
                                                  {(1+n_b) \, m + N b \langle q_m \rangle } ~~,  
\ee

   \noi  
   where $\langle \cdot \rangle$ in \eq{p_n_eksakt} denotes averaging, to be discussed below,  
   with respect to $\theta$ or $\Delta$.  ${\bar p}_0$ is a normalization constant.

      After the passage through the microcavity we make a selective measurement of  
      the atoms. We imagine that one then only measure those atoms  
      leaving the cavity with a definite value of $\theta$ or $\Delta$, i.e.  
      in effect putting a sharp velocity filter or a filter sensitive to detuning at  
      the atom output port of the cavity.  
      The probability ${\cal P}(s)$ of finding an atom in a state   
      $s=\pm$ after the interaction with the cavity photons, where $+$ represents   
      the excited atom state and $-$ represents the atom ground state, can then   
      be expressed in the following matrix form \cite{ElmforsLS95}:

\be  \label{p_pl_matrix}  
  {\cal P}(s) = {\bar u}^{T} M(s) {\bar p} ~~,  
\ee

   \noi  
   such that ${\cal P}(+)\,+\,{\cal P}(-) = 1$.  
   The elements of the vector ${\bar p}$ is given by the equilibrium distribution in \eq{p_n_eksakt}.  
   The quantity ${\bar u}$ is a vector with all entries equal to $1$, ${\bar u}_n=1$.  
   Explicitly ${\cal P}(+)$ takes the form

\be \label{p_pl}  
  {\cal P}(+) = a \sum_{n=0}^{\infty} \; {\bar p}_n ( 1 - q_{n+1}  )   
              + b \sum_{n=0}^{\infty} \; {\bar p}_{n+1} q_{n+1} ~~.  
\ee

    \noi  
    It is well known in the literature that this probability can exhibit  
    quantum revivals (see e.g. Refs. \cite{Walther_div} - \cite{samle_bok_publ}).  
    Furthermore, the joint probability for observing  
    two consecutive atoms in the states $s_1$ and $s_2$ 
    is given by \cite{ElmforsLS95}

\bea  \label{p_PP_M}  
  {\cal P}(s_1,s_2) &=& {\bar u}^{T} S(s_2) \, S(s_1)~ {\bar p} \nonumber \\   
  &=&  
  {\bar u}^{T} M(s_2) \, \, S(s_1)~ {\bar p} ~~,  
\eea

    \noi    
    where $S(s) = (1+L_C/N)^{-1}M(s)$. Explicitly one finds

\bea \label{p_PP}  
  {\cal P}(+,+) &=& a^2 \sum_{n,m=0}^{\infty} \; (1-q_{n+1}) \, (1+L_C/N)^{-1}_{nm} \,   
                (1-q_{m+1}) \, {\bar p}_m \nonumber \\   
                &+&  
                ab \sum_{n,m=0}^{\infty}  \;   
                \Big \{ \, (1-q_{n+1}) \, (1+L_C/N)^{-1}_{nm} \, q_{m+1}   
                \; {\bar p}_{m+1} \nonumber \\  
    &+&  q_n \, (1+L_C/N)^{-1}_{nm} \, (1-q_{m+1})   
                \; {\bar p}_m \, \Big \} \nonumber \\  
                &+& b^2 \sum_{n,m=0}^{\infty} \; q_n \, (1+L_C/N)^{-1}_{nm} \,   
                q_{m+1} \, {\bar p}_{m+1} ~~.  
\eea

    In Refs.\cite{ElmforsLS95,Rekdal99} it is shown that the equilibrium distribution    
    \eq{p_n_eksakt} can be re-written in a form which is rapidly convergent in the   
    large $N$ limit by making use of a Poisson summation technique \cite{Schleich93}.   
    In terms of the scaled photon number variable $x=n/N$ the stationary probability   
    distribution can be written in the from

\be \label{p_x}  
   {\bar p}(x) = {\bar p}_0 \sqrt{\frac{w(x)}{w(0)}} ~ e^{-N \, V(x)}~~,  
\ee

   \noi where   
  
\be  
   V(x) = \sum_{k=-\infty}^{\infty}V_k(x)~~.   
\ee  
  
\noi The effective potential $V(x)$ is expressed in terms of

\be \label{V_k}  
   V_k(x) = -  \int _0^x d\nu  \, \ln[\, w(\nu)\,] \, \cos(2\pi N k \nu) ~~, 
\ee

  \noi where

\be \label{w_x}  
   w(x) = \frac{n_b \, x + a \, \langle q(x) \rangle}{(1+n_b)\, x + b \, \langle q(x) \rangle}  ~~.   
\ee

    \noi  
    We stress that \eq{p_x} is exact.  
    In the large $N$ limit \eq{p_x} can be simplified by making use of a saddle-point  
    approximation. Apart from calculable $1/N$ corrections we put $V(x)=V_0(x)$. The  
    saddle-points are then determined by $V_0^{\prime}(x)=0$, where $V_0^{\prime}(x)$  
    is the derivative of $V_0(x)$ with respect to $x$.  
    Hence, it is the nature of the global minima of $V_0(x)$ which determine   
    the probability distribution, apart from possible zeros of $w(x)$.

   If the only global minimum of $V_0(x)$ occurs at $x=0$, which corresponds to the thermal  
   phase of the micromaser, we can expand   
   the effective potential $V_0(x)$ around origin. The probability ${\bar p}_n$  
   is then given by

\be \label{p_n_geo}  
 {\bar p}_n = {\bar p}_0  
              \left ( \frac{n_b+a\,\theta^{2}_{eff}}{1+n_b+b\,\theta^{2}_{eff}} \right)^n ~~,  
\ee

  \noi  where

\bea \label{theta_eff}  
  \theta_{eff}^2  = \lim_{x \rightarrow 0} \frac{\langle q(x) \rangle}{x}   
  =  \left \langle  \frac{ \sin^2(\theta \Delta)}{\Delta^2} \right \rangle ~~.  
\eea

   \noi  
   The probability distribution  as given by \eq{p_n_geo} is normalizable provided

\be  \label{c_bet}  
   \theta_{eff}^2 (2a-1) < 1 ~~.  
\ee

   If, on the other hand, there exists non-trivial saddle-points of $V_0(x)$,   
   which correspond to the possible maser phases of the micromaser,  
   we write

\be \label{p_x_helt_gen}  
  {\bar p}(x) = \sum_j {\bar p}_j(x) ~~,  
\ee

   \noi  
   where $\sum_j$ denotes the sum over the local minima of $V_0(x)$, i.e. $V_0^{\prime \prime}(x) > 0$,  
   and where ${\bar p}_j(x)$ is ${\bar p}(x)$ for $x$ close to the local    
   minimum $x=x_j$. The distribution ${\bar p}_j(x)$ is given by

\be \label{p_j}  
   {\bar p}_j(x) = \frac{T_j}{\sqrt{2\pi N}}  
                   ~e^{-\frac{N}{2} V_0^{\prime \prime}(x_j)(x-x_j)^2}  ~~,  
\ee

  \noi   
  where $T_j$ is determined by the normalization condition for ${\bar p}(x)$,   
  i.e

\be  \label{T_j}  
   T_j = \frac{e^{- N V_0(x_j)}}{ {\displaystyle \sum_m}   
         \displaystyle{ e^{-N V_0(x_m)}}/\sqrt{V_0^{\prime \prime}(x_m)}}~~,  
\ee

  \noi and

\be \label{V_dobbel}  
   V_0^{\prime \prime}(x) =  \frac{(2a-1)^2}{a + n_b(2a-1)} \, \frac{\langle q(x) \rangle -   
                             x \, d \langle q(x) \rangle/dx }{x^2} ~~.  
\ee

   \noi  
   For the given parameters, the sum in \eq{T_j} is supposed to be taken   
   over all saddle-points corresponding to a minimum of $V_0(x)$,   
   i.e. all saddle-points corresponding to $V_0^{\prime \prime}(x)>0$.  
   If $x=x_j$ does not correspond to a global minimum of the effective potential $V_0(x)$,    
   then $T_j$ is exponentially small in the large $N$ limit.  
   If $x=x_j$ does correspond to one and only one global minimum,  we can neglect all   
   the terms in the sum in $T_j$ but $m=j$, in which case $T_j$ is   
   reduced to  $T_j = \sqrt{V_0^{\prime \prime}(x_j)}$.   
   In the neighbourhood of such a global minimum the probability distribution in   
   \eq{p_j} is therefore reduced to

\be \label{p_j_gauss}  
   {\bar p}_j(x) = \frac{1}{ \sqrt{2\pi N^2 \sigma_x^2}} ~ e^{-\frac{(x-x_j)^2}{2\sigma_x^2}} ~~,  
\ee

    \noi  
    where the standard deviation $\sigma_x$ of \eq{p_j_gauss} is

\be \label{sigma_x}  
  \sigma_x = \frac{1}{ \sqrt{N V_0^{\prime \prime}(x_{j})}}  ~~,  
\ee

  \noi  
  which is zero in the large $N$ limit provided $V_0^{\prime \prime}(x_j) \neq 0$.   
  The probability ${\bar p}(x)$ is therefore peaked around the global minima for large $N$.  
  The minimum $x_j$ of the potential $V_0(x)$ is a solution of saddle-point equation

\be \label{sadel_lign}  
  x = (2a-1) \langle q(x) \rangle \leq 1 ~~.  
\ee  
  
\noi This saddle-point equation is independent of the number $n_b$ of thermal photons in the cavity.

    The excitation probability ${\cal P}(+)$ can be re-written in a form where quantum revivals   
    (see e.g. Ref. \cite{Knight93}) become explicitly by making use of   
    a Poisson summation technique \cite{Schleich93}. When $a=1$ we obtain

\bea \label{p_P_Poisson_gen}  
  {\cal P}(+) &=&  1 - \frac{1}{2} \sum_{n=0}^{\infty} \, {\bar p}_n \frac{n+1}{n+1+ N \Delta^2} +  
              \frac{1}{2} \sum_{\nu=-\infty}^{\infty} w_{\nu}(\theta)   
              \nonumber \\  
              &+& \frac{1}{2} \, {\bar p}_0 \; \frac{1}{1+ N \Delta^2}    
              \cos( \, 2 \theta \sqrt{1/N + \Delta^2}\, ) ~~,  
\eea

   \noi  
   where

\be \label{w_0_gen}  
    w_0(\theta) = N \int_0^{\infty} dx \; {\bar p}(x)
    \frac{x+1/N}{x+1/N + \Delta^2}
\cos \left( 2 \theta    
    \sqrt{x + 1/N  + \Delta^2 } \right ) ~~,  
\ee

  \noi and

\bea \label{w_nu_gen}  
   w_{\nu}(\theta) &=& N \; \mbox{Re} \,  
                     \bigg \{ \; e^{-2\pi i \nu N \left[ \,\left ( \, \theta/2 \pi \nu N \,   
                     \right )^2  +  \Delta^2 \, \right ] } \nonumber \\ &\times&   
                     \int_0^{\infty} dx \; {\bar p}(x) \, \frac{x+1/N}{x+1/N + \Delta^2} \;   
                     e^{2\pi i \nu N \left( \, \sqrt{x+1/N + \Delta^2} -   
                     \theta/2 \pi \nu N \, \right )^2  } \; \bigg \} ~~.  
\eea

     \noi  
     Here ${\bar p}(x)$ is the continuous version of ${\bar p}(n/N)$.   
     We stress that \eq{p_P_Poisson_gen} is exact. If $a\neq 1$ we
     obtain  ${\cal P}(+)$ by the replacement ${\cal P}(+) \rightarrow
     1-a + (2a-1){\cal P}(+)$, apart from $1/N$ corrections.
     If the probability distribution ${\bar p}(x)$ is sufficiently peaked, i.e. ${\bar x} \gg \sigma_x$,  
     where ${\bar x} \equiv {\bar x}(\theta)$ denotes the average of $x=n/N$ with respect to the   
     stationary probability distribution ${\bar p}_n$, then the excitation probability is reduced to

\be \label{p_P_Poisson}  
  {\cal P}(+) \approx 1 - \frac{1}{2} \frac{{\bar x}}{ {\bar x} + \Delta^2}   
              + \frac{1}{2} \sum_{\nu=0}^{\infty} w_{\nu}(\theta) ~~,  
\ee

  \noi  
  where

\bea \label{w_nu}  
  w_{\nu}(\theta) &\approx& {\bar p}( x = {\bar x}_{\nu}) \,  
                  \frac{{\bar x}}{ {\bar x} + \Delta^2}\,   
                  \frac{\theta}{\pi \sqrt{2 \nu^3 N}} \;   
                  \nonumber \\ &\times&  
                  \cos \left [ \, 2 \pi \nu  N \left ( \, \left ( \, \frac{\theta}{2 \pi \nu N} \,   
                  \right )^2 + \Delta^2 \, \right )  - \frac{\pi}{4} \, \right ] ~~, ~~ \nu \geq 1 ~~,  
\eea

    \noi  and

\be \label{n_nu_2}  
  {\bar x}_{\nu} = \left ( \, \frac{\theta}{2 \pi \nu N} \, \right )^2 - \Delta^2 ~~.  
\ee

   \noi  
   According to \eq{n_nu_2} the $\nu$:th revival of ${\cal P}(+)$ occurs in the region   
   where $\theta$ is close to

\be \label{overl_theta_nu}  
   \theta_{\nu} = 2 \pi \nu N \sqrt{ {\bar x}  + \Delta^2} ~~.  
\ee

     \noi  
     This equation relates the width $\sigma_x$  of the probability distribution ${\bar p}(x)$  
     into a measure for the width $\Delta \theta_{\nu}$ in the pump parameter of the $\nu$:th revival  
     according to a probability distribution of the form   
        $p(\theta) \simeq \exp(-(\theta -\theta_{\nu})^2/2(\Delta \theta_{\nu})^2)$, i.e.

\be \label{sig_mu}  
  \Delta \theta_{\nu} =  \pi \nu N \,  \frac{\sigma_x}{\sqrt{ {\bar x} + \Delta^2} } ~~.  
\ee

    \noi  
    Two consecutive terms of the sum in \eq{p_P_Poisson} separate in time when their   
    temporal separation $\theta_{\nu+1} - \theta_{\nu}$ is larger then  
    $\Delta \theta_{\nu+1} + \Delta \theta_{\nu}$, i.e. within one standard deviation,  
   provided

\be \label{ikke_overlapp_bet_NY}  
  \nu < \frac{ {\bar x} + \Delta^2 }{ \sigma_x} - \frac{1}{2} ~~.  
\ee

    \noi  
    The revivals in ${\cal P}(+)$ cannot be resolved for values of $\nu$ larger then the right-hand   
    side of \eq{ikke_overlapp_bet_NY}.

    If the probability distribution ${\bar p}(x)$ is sufficiently
    peaked, 
    i.e. ${\bar x} \gg \sigma_x$,  
    then the excitation probability ${\cal P}(+)$ approaches   
  
\be \label{P_a_peaked}  
  {\cal P}(+) = a- \frac{1}{2}(2a-1)\left( \frac{{\bar x}}{({\bar x} + 
\Delta^2)} - w_{0}(\theta)\right) \approx a -(2a-1)q(x)~~,  
\ee

    \noi  
    in the large $N$ limit, where we make use of

\be
\label{eq:asymp}
    w_{0}(\theta) \approx 
\exp(-\theta^2 /2N)\frac{\bar x}{{\bar x}+\Delta^2}\cos(2\theta\sqrt{{\bar
    x}+\Delta^2}) ~~.
\ee
If the average in the saddle-point equation
    \eq{sadel_lign} is such that $\langle q(x) \rangle \approx q(x)$,
    we see that ${\cal P}(+) \approx a - {\bar x}$. If we average
    \eq{eq:asymp} with respect to the noise in the system, we obtain 
  $\langle {\cal P}(+) \rangle \approx a - {\bar x}$ by making use of
  the
saddle-point equation \eq{sadel_lign} once more. 

The approximative result \eq{P_a_peaked} can actually be converted
     into an exact and more general relation between ${\bar x}$ and ${\cal P}(+)$.
      If the atoms are not measured with well-defined $\theta$ or
      $\Delta$ then we must in general  
      average ${\cal P}(+)$ with respect to the corresponding probability distribution.  
      In \eq{p_pl} we then replace $q_n$ by $\langle q_n \rangle$ and denotes the corresponding  
      excitation probability ${\cal P}(+)$ by $\langle {\cal P}(+) \rangle$.   
      From the equilibrium distribution \eq{p_n_eksakt} we then obtain   
      the recurrence formula   
      $[ \, (1+n_b) \, n + N b \langle q_n \rangle \, ] \, 
       {\bar p}_n = {\bar p}_{n-1} [ \, n_b \, n +    
      N a \langle q_n \rangle \, ]$. By summing this formula over $n$, we therefore derive that

\be \label{sh_xP}  
   {\bar x} = a + \frac{n_b}{N} - \langle {\cal P}(+) \rangle ~~.  
\ee

     \noi   
     In general and in the presence of noise in $\theta$ and/or $\Delta$, ${\bar x}$ and ${\cal P}(+)$   
     have no direct relation and we must in general consider them to be independent variables.

\vspace{0.5cm}  
  
\bc{  
\section{Absence of Fluctuations}  
\label{no_avg_SEC}  
}\ec  
  
\vspace{0.5cm}

\begin{figure}[t]  
\unitlength=1mm  
\begin{picture}(100,80)(0,0)

\includegraphics{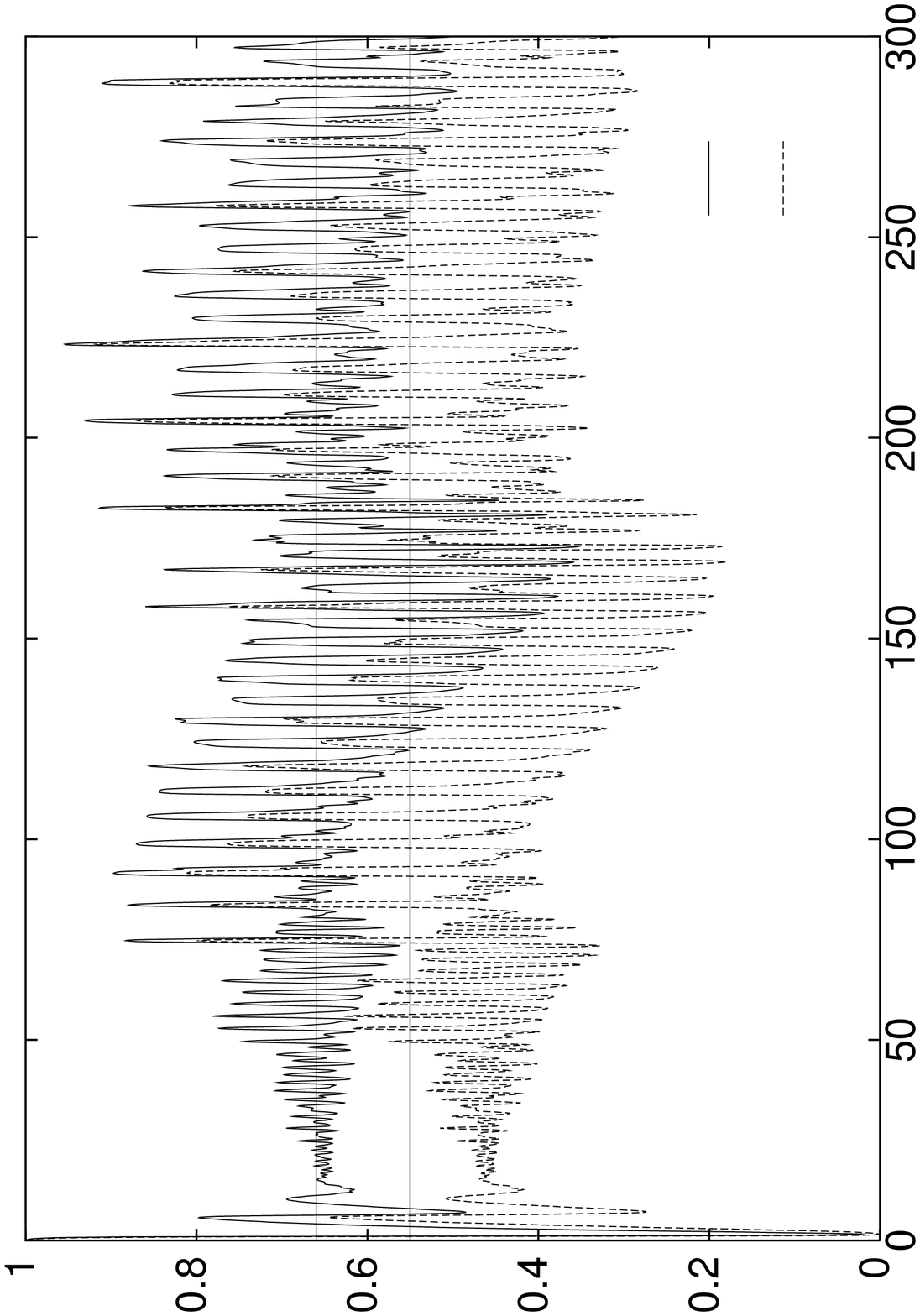}                                        

   \put(79,-6){\normalsize     \boldmath$\theta$}  
   \put(113.5,22.5){\normalsize  \boldmath${\cal P}(+)$}  
   \put(108,14){\normalsize  \boldmath${\cal P}(+,+)$}

\end{picture}  
\vspace{8mm}  
\figcap{ The probabilities ${\cal P}(+)$ (solid line) and ${\cal P}(+,+)$ (dotted line) as a   
         function of $\theta$ when the $q_m$-terms in \eq{p_n_eksakt} are not averaged.   
         The parameters are: $a=1$, $\Delta=0$, $n_b=0.15$ and $N=35$.  
         The curves show no clear evidence for the resonant behavior of revivals.   
         In the large $\theta$ and $N$ limit the probability ${\cal P}(+)$ approaches   
         ${\cal P}_{\infty}(+) \approx 0.66$ and the probability ${\cal P}(+,+)$ approaches   
         ${\cal P}_{\infty}(+,+) \approx 0.57$, 
         which are indicated by solid lines in the figure.  
\label{p_pl_sig0_THETA_FIG} }  
\end{figure}

     If the atoms enter the cavity without a spread in the transit time   
     and without a spread in the atom-photon frequency detuning,   
     the behavior of the order parameter and the excitation probability  
     ${\cal P}(+)$ is well known in the literature   
     (see e.g. Refs. \cite{Walther_div,samle_bok_publ,Filipowicz86_igjen}).  
     As e.g. seen in \fig{p_pl_sig0_THETA_FIG}, ${\cal P}(+)$ under such circumstances shows no  
     clear evidence for resonant behavior of revivals. A more precise way to express the presence of  
     revivals in  
     the temporal variations of ${\cal P}(+)$ can e.g. obtained by performing  
     a Fourier transformation of ${\cal P}(+)$ and considering the width of the corresponding spectrum.   
     The corresponding Fourier spectrum  
     of \fig{p_pl_sig0_THETA_FIG} is then broad. The appearance of clear revivals   
     would correspond to a more peaked Fourier spectrum.

     In the large $N$ and $\theta$ limit the order parameter approaches a constant. This constant,  
     ${\bar x}_{\infty}$, is therefore determined by the   
     global minimum of $V_0(x)$ in the large $\theta$ limit, in which case $V_0(x)$ has a unique minimum.  
     The micromaser system has then no maser-maser phase transitions.  
     As shown in an Appendix, this minimum is determined  by    
     $\partial V_0(x)/\partial x=0$, where

\be \label{hurra_formel}  
     \frac{\partial V_0(x)}{\partial x} = \ln \left [ \frac{1-a}{a} \right ] +   
     f(\, y(x) \, ) - f( \, w(x) \,) ~~,  
\ee

    \noi   
    and where $f(z)$, $y(x)$ and $w(x)$ are as given in the Appendix.  
    When e.g. $a=1$, $n_b=0.15$ and $\Delta=0$ the global minimum of $V_0(x)$ occurs at   
    ${\bar x}_{\infty} \approx 0.34$, which corresponds to an asymptotic excitation probability   
    ${\cal P}_{\infty}(+) \approx 0.66$ due to \eq{sh_xP}.   
    When $a=1$ and $\Delta=0$ it follows from \eq{p_PP} that for large $\theta$  
    the joint probability ${\cal P}(+,+)$ approaches the asymptotic value

\be \label{P_pp_a_smh}  
  {\cal P}_{\infty}(+,+) \approx (5 {\cal P}_{\infty}(+) - 1)/4 ~~,  
\ee

    \noi  
    i.e. with the physical parameters as above we find ${\cal P}_{\infty}(+,+) \approx 0.57$   
    (see \fig{p_pl_sig0_THETA_FIG}). \eq{P_pp_a_smh} follows from \eq{p_PP} by making use of  
    the following properties of the unbounded operator $L_C$

\be  
  \sum_{n=0}^{\infty} \, (L_C)_{nm} = 0  ~~ , ~~  \sum_{m=0}^{\infty} \, (L_C)_{nm} = -1  ~~,  
\ee

    \noi  
    and performing a suitable large $N$ limit.

\vspace{1.0cm}  
  
\bc{  
\section{Effects of Fluctuations}  
\label{fluctuations_SEC}  
}\ec  
  
\vspace{0.0cm}


      We will now consider physical effects of noise in the micromaser system.  
      We start by discussing the effects of adding fluctuations to the pump parameter $\theta$ and  
      then, in Section \ref{det_SEC}, we study the effects of adding fluctuations to the  
      atom-photon detuning parameter $\Delta$.

\vspace{0.5cm}  
  
\bc{  
\subsection{Pump Parameter Fluctuations}  
\label{pump_SEC}  
}\ec  
  
\vspace{0.0cm}

     Suppose that the pump parameter $\theta$ is described by a positive stochastic variable $\xi$   
     as described  
     by the probability distribution $P_{\theta}(\xi)$ such that $\langle \xi \rangle = \theta$ and   
     $\langle ( \xi - \theta )^2 \rangle = \sigma_{\theta}^2$. The averaged value of $q(x)$  
     with respect to $P_{\theta}(\xi)$ is then given by

\be \label{q_t_def}  
  \langle q(x) \rangle_{\theta} = \int_{0}^{\infty} d \xi \;   
   P_{\theta}(\xi) \, \frac{x}{x+\Delta^2} \sin^2 \left ( \xi \sqrt{x+\Delta^2} \right )  ~~.  
\ee

    \noi  
    Non-trivial saddle-points of the corresponding $V_0(x)$ can be found by solving the equation

\be \label{s_t}  
   1 = (2a-1) \, I_1(\theta,x(\theta)) ~~,  
\ee

  \noi where

\be \label{I_1_def_gen}  
    I_1(\theta, x) = \frac{1}{x+\Delta^2}\; {\int_{0}^{\infty} d \xi \; P_{\theta}(\xi) \,   
    \sin^2 \left ( \, \xi  \sqrt{x + \Delta^2} \; \right ) \leq \sigma_{\theta}^2 + \theta^2}  ~~.  
\ee

     For small values of $x$, we can expand the effective potential $V_0(x)$ around $x=0$.  
     A straightforward expansion $V_0(x)$ then leads to

\bea \label{V_exp_fase}  
   V_0(x) &=& x \ln \left [ \frac{1 + n_b+a\,\theta^{2}_{eff}}{n_b+a\,\theta^{2}_{eff}} \right ]  
    \nonumber \\ &+&   
    \frac{x^2}{2}   \frac{a+n_b(2a-1)}{(n_b + a \theta_{eff}^2)(1+n_b+b \theta_{eff}^2)} ~  
    \langle \, \xi^4 f(\xi |\Delta| ) \, \rangle  \; + \; {\cal O}(x^3)~~,  
\eea

    \noi  
    where   
  
\be  
   f(x) = \frac{\sin^2(x)}{x^4} - \frac{\sin(x) \cos(x)}{x^3} ~~.  
\ee

     \noi  
     For small $x$, i.e. $x \ll 1$, $f(x) = 1/3 + {\cal O}(x^2)$.   
     For typical physical parameters such that $\theta |\Delta| \lesssim \pi$, we therefore  
     observe a thermal-maser phase transition for values of $\theta$ determined by $V_0^{\prime}(x)=0$.  
     The corresponding critical transition line is then

\be \label{a_t_gen}  
  a(\theta) = \frac{1}{2} + \frac{1}{2} \frac{1}{I_1(\theta)} \leq 1  ~~,  
\ee

     \noi  
     where $I_1(\theta,0) \equiv I_1(\theta)$. \eq{a_t_gen} coincides with the radius of convergence   
     of the thermal probability distribution \eq{p_n_geo} as it should.  
     For $\Delta=0$ we see that $\theta_{eff}^2 = \sigma_{\theta}^2 + \theta^2$ and  
     hence this phase transition can occur for $\theta=0$, i.e. the phase transition   
     can be induced by the presence of noise in the stochastic variable $\theta$.   
     For $\theta |\Delta| \gg \pi$ we must in general make use of a combination   
     of analytical and numerical methods, as in the case of $\sigma_{\theta}=0$ \cite{Rekdal99},  
     in order to get a detailed picture of the phase diagram.

   In the maser phase,   
   the order parameter ${\bar x}(\theta)$ approaches zero when the system 
   approaches the critical line \eq{a_t_gen}. Furthermore,  
   in the large $N$ limit ${\bar x}(\theta)$ is always zero in the thermal phase.   
   Hence, the order parameter is continues on the critical transition line \eq{a_t_gen}.  
   To determine the order of the phase transition we therefore have to investigate   
   higher order derivatives.  
   The first derivative of ${\bar x}(\theta)$ with respect to $\theta$, ${\bar x}^{\prime}(\theta)$,  
   at the critical transition line \eq{a_t_gen} is

\be  \label{dx_dtheta}  
  {\bar x}^{\prime}(\theta) = \frac{\Delta^2 \, I_1^{\prime}(\theta)}  
                        {I_1(\theta) - I_2(\theta)} ~~,  
\ee

   \noi  
   where  $I_1^{\prime}(\theta)$ is the derivative of $I_1(\theta)$ with respect to $\theta$  and     
   where

\be \label{I_2}  
     I_2(\theta)  = \int_{0}^{\infty} d \xi \;   
     P_{\theta}(\xi) \, \xi^2  \; \frac{\sin \left ( 2 \xi \Delta \right )}{2\xi \Delta}  ~~.  
\ee

     \noi  
     By making use of the positive definiteness of $P_{\theta}(\xi) \xi^2$, the  
     Cauchy-Schwarz inequality implies  
     $|I_2(\theta)| \leq \sqrt{\sigma_{\theta}^2 + \theta^2}$.  
     If $x^{\prime}(\theta)$ is zero, then we have to  
     investigate higher order derivatives. The second derivative of ${\bar x}(\theta)$   
     with respect to $\theta$,  ${\bar x}^{\prime \prime}(\theta)$, at the critical   
     thermal-maser line \eq{a_t_gen} is

\be  
   {\bar x}^{\prime \, \prime}(\theta) = \frac{\Delta^2 I_1^{\prime \, \prime}(\theta) +   
   {\bar x}^{\prime}(\theta) I_2^{\prime}(\theta)}{I_1(\theta) - I_2(\theta)} ~~,  
\ee

     \noi  
     where $I_1^{\prime \, \prime}(\theta)$ is the second derivative of $I_1(\theta)$ with respect   
     to $\theta$ and $I_2^{\prime}(\theta)$ is the derivative of $I_2(\theta)$ with respect to $\theta$.

   The effective potential $V_0(x)$ approaches its large $\theta$ limit exponentially fast     
   as a function of $\sigma_{\theta}^2$.  
   This unique minimum is determined by $\partial V_0(x)/\partial x=0$, where   
   $\partial V_0(x)/\partial x$ is given by \eq{hurra_formel} with $y(x)$ and $w(x)$ replaced   
   by ${\tilde y}(x) = a \, e^{-2(x+\Delta^2) \sigma_{\theta}^2}/[ \,     
   \frac{a}{2} (1+e^{-2(x+\Delta^2) \sigma_{\theta}^2} ) +n_b( x+\Delta^2 ) \,  ]$ and   
   ${\tilde w}(x) = b \, e^{-2(x+\Delta^2) \sigma_{\theta}^2}/[ \,     
   \frac{b}{2} (1+e^{-2(x+\Delta^2) \sigma_{\theta}^2} ) +(1+n_b)( x+\Delta^2 ) \,  ]$,  
   respectively.

    If the micromaser is not detuned, i.e. $\Delta=0$, then there is one   
    critical thermal-maser transition line only. This critical transition line  
    can be found analytically for any probability distribution $P_{\theta}(\xi)$.  
    The integral $I_1(\theta)$ is then given by

\be  
   I_1(\theta) = \sigma_{\theta}^2 + \theta^2 \geq 1 ~~.  
\ee

    \noi  
    The critical thermal-maser transition line is then only dependent   
    of the variance $\sigma_{\theta}^2$ and the mean value $\theta$:

\be \label{t_m_line}  
  a(\theta) = \frac{1}{2} + \frac{1}{2} \frac{1}{\sigma_{\theta}^2  + \theta^2 } ~~.  
\ee

\begin{figure}[t]  
\unitlength=1mm  
\begin{picture}(100,80)(0,0)

\includegraphics{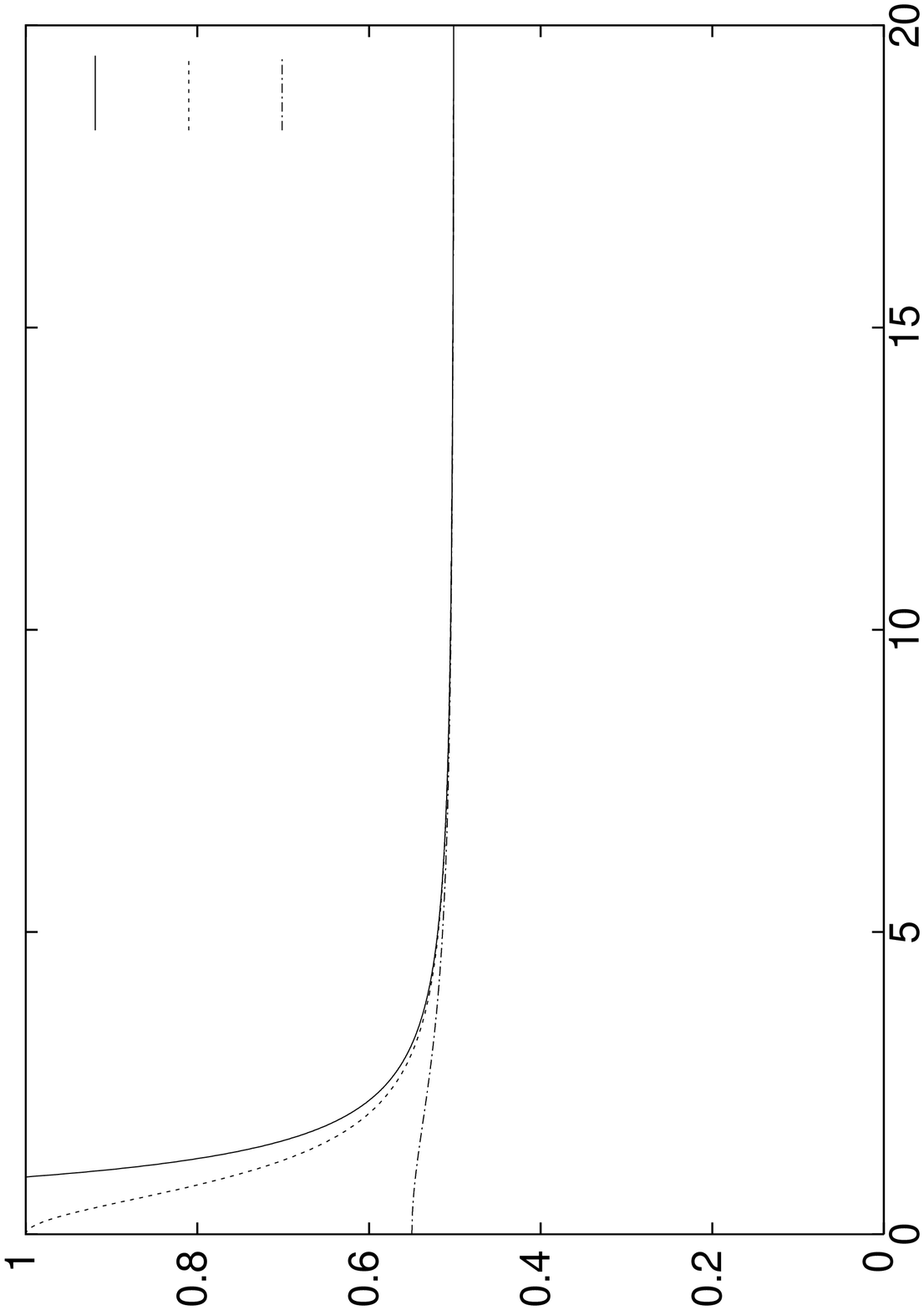}

   \put(2,52){\normalsize     \boldmath$a$}  
  
   \put(80,-6){\normalsize     \boldmath$\theta$}  
   \put(115,92.5){\normalsize  \boldmath$\sigma_{\theta}^2=0.1$}  
   \put(118.4,82){\normalsize  \boldmath$\sigma_{\theta}^2=1$}  
   \put(116,71){\normalsize    \boldmath$\sigma_{\theta}^2=10$}

\end{picture}  
\vspace{8mm}  
\figcap{The thermal-maser critical lines $a(\theta)$ according to \eq{t_m_line} for   
        $\sigma_{\theta}^2=0.1,1,10$, i.e. when the micromaser is not detuned. All other parameters   
        are as in Fig. \ref{p_pl_sig0_THETA_FIG}. When $\sigma_{\theta}^2 \geq 1$ the critical   
        thermal-maser line describes a third-order  
        phase transition at $a(0)=1/2+1/(2\sigma_{\theta}^2)$ and a second-order transition for any  
        other possible value of $a$. When $\sigma_{\theta}^2=10$, this third-order phase transition  
        occurs at $a(0)=0.55$.   
\label{fase_diff_sig_FIG} }  
\end{figure}

   \noi  
   For values of $a$ and $\theta$ above this critical line, i.e. $a \geq a(\theta)$, the micromaser  
   system is in a maser phase. Below this line, on the other hand, the micromaser is   
   in the thermal phase (see e.g. \fig{fase_diff_sig_FIG}).

    The order parameter ${\bar x}(\theta)$ is continuous on the critical line \eq{t_m_line}.  
    The first derivative of ${\bar x}(\theta)$ with respect to $\theta$, ${\bar x}^{\prime}(\theta)$,  
    at this critical thermal-maser line is

\be   
    {\bar x}^{\prime}(\theta) = 6 \, \frac{\theta}{\langle \xi^4 \rangle} ~~,  
\ee

    \noi  
    which is non-zero for $\sigma_{\theta}^2 < 1$. The critical line   
    \eq{t_m_line} then describes a line of second-order (thermal-maser) phase transition.   
    If $\sigma_{\theta}^2 \geq 1$,  
    \eq{t_m_line} also describes a line of second-order phase transitions for all values of $\theta$  
    except for $\theta=0$ where $a(0)=1/2 + 1/(2 \sigma_{\theta}^2)$. Since ${\bar x}^{\prime}(\theta)$   
    then is zero we have to investigate higher order derivatives. The second derivative of ${\bar x}(\theta)$   
    with respect to $\theta$, ${\bar x}^{\prime \prime}(\theta)$, at the critical thermal-maser line is

\be \label{dobbel_x}  
   {\bar x}^{\prime \, \prime}(\theta)  
   = \frac{6}{\langle \xi^4 \rangle}  
        \left [ \, 1 -  2 \frac{\theta}{\langle \xi^4 \rangle}  
        \frac{d \langle \xi^4 \rangle}{d \theta} + \frac{7}{12}   
        \left (\frac{ \theta }{ \langle \xi^4 \rangle  } \right )^2  \langle \xi^6 \rangle \, \right ] ~~,  
\ee

   \noi  
   which is non-zero when the first derivative ${\bar x}^{\prime}(\theta)$ is zero. The point   
   $a(0)=1/2 + 1/(2 \sigma_{\theta}^2)$ on the critical thermal-maser line therefore corresponds to  
   a third-order transition. From \eq{dobbel_x} it follows, in addition, that we can have at most a   
   third-order phase transition when fluctuations in $\theta$ are taken into account.

        If the probability $P_{\theta}(\xi)$ is sufficiently peaked such that the   
        its $\xi$-variation is fast in comparison to $\xi$-variation of   
        $\sin^2 \left ( \xi \Delta \right )$, i.e. if $\sigma_{\theta} |\Delta| \ll \pi/2$,  
        then \eq{I_1_def_gen} is reduced to  
        $I_1(\theta) = \sin^2 \left ( \theta \Delta \, \right )/\Delta^2 + \sigma_{\theta}^2  \,   
        \cos \left ( 2 \theta \Delta \, \right )$.   
        The first thermal-maser critical line is then given by

\be  \label{a_kons}  
  a(\theta) = \frac{1}{2} + \frac{1}{2} \frac{\Delta^2}{\sin^2(\theta \Delta) + \sigma_{\theta}^2  
              \Delta^2 \cos( 2 \theta \Delta )} ~~,  
\ee

     \noi   
     which is consistent with the result of Ref. \cite{Rekdal99} for $\sigma_{\theta}^2=0$.

    If, on the other  hand, the quantity $P_{\theta}(\xi)$ has a   
    $\xi$-variation which is slow in comparison to the $\xi$-variation of   
    $\sin^2 \left ( \xi \Delta \right )$   
    i.e. if $\sigma_{\theta} |\Delta| \gg \pi/2$, then  \eq{I_1_def_gen}  
    reduces to $I_1(\theta) = 1/2\Delta^2$.  
    \eq{a_t_gen} is then reduced to

\be  
  a(\theta) = \frac{1}{2} + \Delta^2 ~~.  
\ee

     In order to get explicit analytic results we now choose the following gamma probability  
     distribution for the pump parameter

\be  
   P_{\theta}(\xi) = \frac{\beta}{\Gamma(\alpha + 1)}  \,  \xi^{\alpha} \, e^{-\beta \xi}  ~~,    
\ee

    \noi  
    where $\beta = \theta/\sigma_{\theta}^2$ and $\alpha = \theta^2/\sigma_{\theta}^2 - 1$, so that  
    $\langle \xi \rangle = \theta$ and $\langle (\xi - \theta)^2 \rangle = \sigma_{\theta}^2$.  
    Other choices are possible, but are not, as long as the distribution is sufficiently peaked,  
    expected to change the overall qualitative picture.      
    The integral $I_1(\theta,x)$ is then given by

\bea \label{I_1_expl}  
    I_1(\theta,x)  &=& \frac{1}{2} \, \frac{1}{x+\Delta^2}   
    \nonumber \\ &\times&  
    \left [ ~ 1 - \bigg( 1 + \frac{2(x+\Delta^2)   
                      \sigma_{\theta}^2}{\theta^2/2 \sigma_{\theta}^2  } \bigg)  
                                     ^{- \frac{\theta^2}{2 \sigma_{\theta}^2} }   
                                     \cos \left(  \, \frac{\theta^2}{\sigma_{\theta}^2}   
                                     \arctan \left( \sqrt { \frac{2(x+\Delta^2)   
                                                    \sigma_{\theta}^2}{\theta^2/2 \sigma_{\theta}^2  }  }  
                                     \right ) \, \right ) ~ \right ] . ~~~~~~  
\eea

   \noi  
   This averaged form of $q(x)$ depends on the two independent variables   
   $\theta^2/(2\sigma_{\theta}^2)$ and $2( x+\Delta^2 ) \, \sigma_{\theta}^2$ only.  
   In particular, if

\be \label{c_theta}  
   \frac{\theta^2}{2 \sigma_{\theta}^2} \gg \sigma_{\theta}^2 ~~,    
\ee

   \noi  
   then  \eq{I_1_expl} reduces to

\be  \label{I_1_red}  
  I_1(\theta,x) =  \frac{1}{2} \, \frac{1}{x+\Delta^2} \;   
  \bigg [ ~ 1 - e^{-2(x+\Delta^2) \sigma_{\theta}^2 } \,  
  \cos\left ( 2 \theta \sqrt{x+\Delta^2} \, \right ) ~ \bigg ] ~~.  
\ee

      \noi  
      This explicit form of $I_1(\theta,x)$ can be achieved for a broad class of probability   
      distributions, e.g. a Gaussian distribution. If, in addition,

\be  \label{bet_stor_theta}  
   \sigma_{\theta}^2 \gg \frac{1}{2a-1} ~~,  
\ee

    \noi  
    then any solution to the corresponding saddle-point equation   
    $1=(2a-1) I_1(\theta, x(\theta) )$ is exponentially close to

\be  \label{num_5_x}  
  {\bar x} = a-1/2 - \Delta^2 ~~.  
\ee

     \noi    
     If, on the other hand,

\be \label{bet_liten}  
  \sigma_{\theta}^2 \gg \theta \gtrsim \sigma_{\theta} \gtrsim 1 ~~,   
\ee

\begin{figure}[t]  
\unitlength=1mm  
\begin{picture}(100,80)(0,0)  
  
\includegraphics{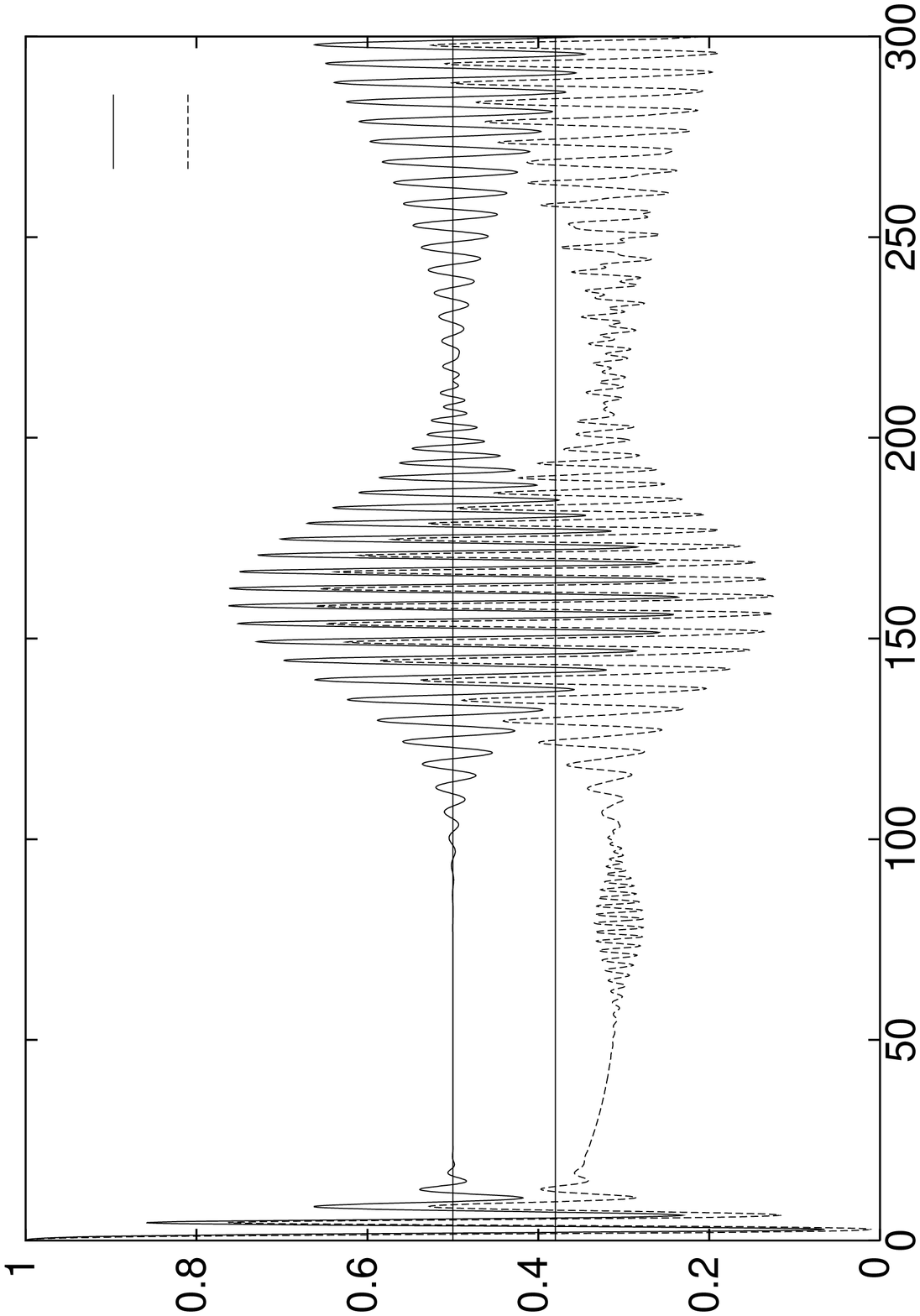}        
  
   \put(80,-6){\normalsize  \boldmath$\theta$}  
   
   \put(120,90.7){\normalsize  \boldmath${\cal P}(+)$}  
   \put(114,82){\normalsize  \boldmath${\cal P}(+,+)$}

\end{picture}  
\vspace{8mm}  
\figcap{ The probabilities ${\cal P}(+)$ (solid line) and ${\cal P}(+,+)$ (dotted line) as a   
         function of $\theta$ when the $q_m$-terms in \eq{p_n_eksakt} are averaged with respect to  
         $\theta$ with $\sigma_{\theta}^2=25$.   
         All other parameters are as in \fig{p_pl_sig0_THETA_FIG}.  
         In the large $\theta$ limit the probability ${\cal P}(+)$ approaches   
         ${\cal P}_{\infty}(+) = 0.5$   
         and the probability ${\cal P}(+,+)$ approaches ${\cal P}_{\infty}(+,+) \approx 0.38$,   
         which are indicated by solid lines in the figure.  
\label{p_pl_sig25_THETA_FIG} }  
\end{figure}

\begin{figure}[t]  
\unitlength=1mm  
\begin{picture}(100,80)(0,0)  
  
\includegraphics{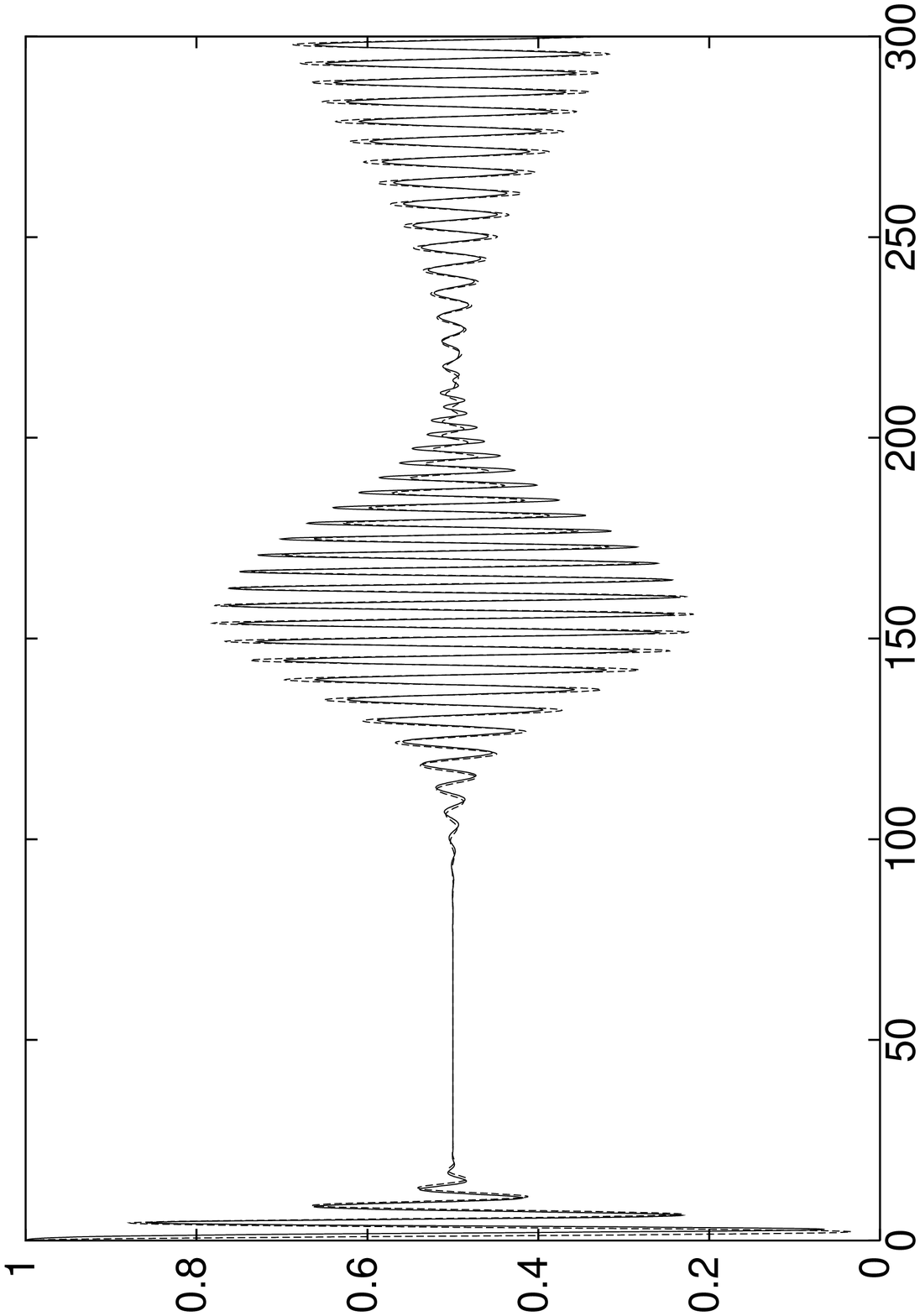}        
  
   \put(80,-6){\normalsize  \boldmath$\theta$}  
   \put(-5,53){\normalsize  \boldmath${\cal P}(+)$}

\end{picture}  
\vspace{8mm}  
\figcap{ Comparison of the exact ${\cal P}(+)$ (solid line) and ${\cal P}(+)$ as given   
         in \eq{p_P_Poisson} (dashed line) with ${\bar p}(x)$ as given in \eq{p_j_gauss}  
         with ${\bar x} = 1/2$.  
         The parameters are the same as in \fig{p_pl_sig25_THETA_FIG}. In this figure  
         we only see the first two revivals, i.e. only the terms $\nu =0,1,2$ in    
         \eq{p_P_Poisson} is visible in this figure.   
\label{p_pl_sig25_SmL_FIG} }  
\end{figure}

     \noi  
     then the corresponding saddle-point equation has one non-trivial solution only, i.e. \eq{num_5_x}.  
     Under the conditions given by Eqs. (\ref{c_theta}) and (\ref{bet_stor_theta}) or \eq{bet_liten}  
     the maser-maser phase transitions will be less significant.  
     The phase diagram then essentially consists of critical thermal-maser line(s) only  
     (see e.g. \fig{fase_diff_sig_FIG}).   
     With regard to the observable ${\cal P}(+)$, the requirement $\theta \gtrsim \theta_{\nu=1}  
     = 2 \pi N \sqrt{{\bar x}+\Delta^2}$ is compatible with \eq{c_theta} provided that $N$ is  
     sufficiently large, i.e.

\be  
\label{verylargen}  
   N \gg \frac{\sqrt{2}}{2 \pi} \, \frac{\sigma_{\theta}^2}{\sqrt{a-1/2}}  ~~,  
\ee

     \noi  
     for the saddle-point solution \eq{num_5_x}.  
     The equilibrium probability distribution is in this case given by \eq{p_j_gauss}  
     with the variance  $\sigma_x^2 = [ \, a + n_b(2a-1) \, ] /2N$ and $x_j={\bar x}$ as in \eq{num_5_x}.  
     The equilibrium distribution ${\bar p}(x)$ is peaked around ${\bar x}$ provided    
     ${\bar x} \gg \sigma_x$, i.e.

\be \label{N_sig_bet_atter}  
   N \gg \frac{1}{2} \; \frac{a + n_b(2a-1)}{[ \, a-1/2 - \Delta^2 \, ]^2} ~~.  
\ee

   \noi  
   From \eq{overl_theta_nu} we observe that the $\nu$:th revival   
   in ${\cal P}(+)$ occurs in the region where $\theta$ is close to

\be  
  \theta_{\nu} = 2 \pi \nu N \sqrt{ a-1/2 } ~~.  
\ee

    \noi  
    According to \eq{ikke_overlapp_bet_NY}, two consecutive revivals in ${\cal P}(+)$   
    are separated provided

\be \label{ikke_overlapp_bet_NY_gauss}  
  \nu < \sqrt{2 N} \; \frac{2a-1}{\sqrt{a+n_b(2a-1)}}  - \frac{1}{2} ~~,  
\ee

      \noi  
      which gives a bound on how many revivals that can be resolved in the excitation probability  
      ${\cal P}(+)$.  
      In Fig. \ref{p_pl_sig25_THETA_FIG}  we consider values of
      relevant physical  parameters such that  
      Eqs. (\ref{bet_stor_theta}), (\ref{verylargen}) and (\ref{N_sig_bet_atter}) are satisfied.   
      We then observe that ${\cal P}(+)$ and ${\cal P}(+,+)$  
      exhibit well pronounced revivals. For completeness we show in Fig. \ref{p_pl_sig25_SmL_FIG},   
      for the same set of values of the physical parameters as in Fig. \ref{p_pl_sig25_THETA_FIG},   
      that the approximative but analytical expression for ${\cal
      P}(+)$ as given  by \eq{p_P_Poisson}  is in very good
      agreement with  the exact result for ${\cal P}(+)$. Within one standard  
      deviation we expect, according to \eq{ikke_overlapp_bet_NY_gauss}, that at most seven 
      revivals in ${\cal P}(+)$ are well separated in this case. 
      Numerically we find that actually only the first   
      two revivals are well separated.

\begin{figure}[p]  
\unitlength=1mm  
\begin{picture}(100,80)(0,0)

\includegraphics{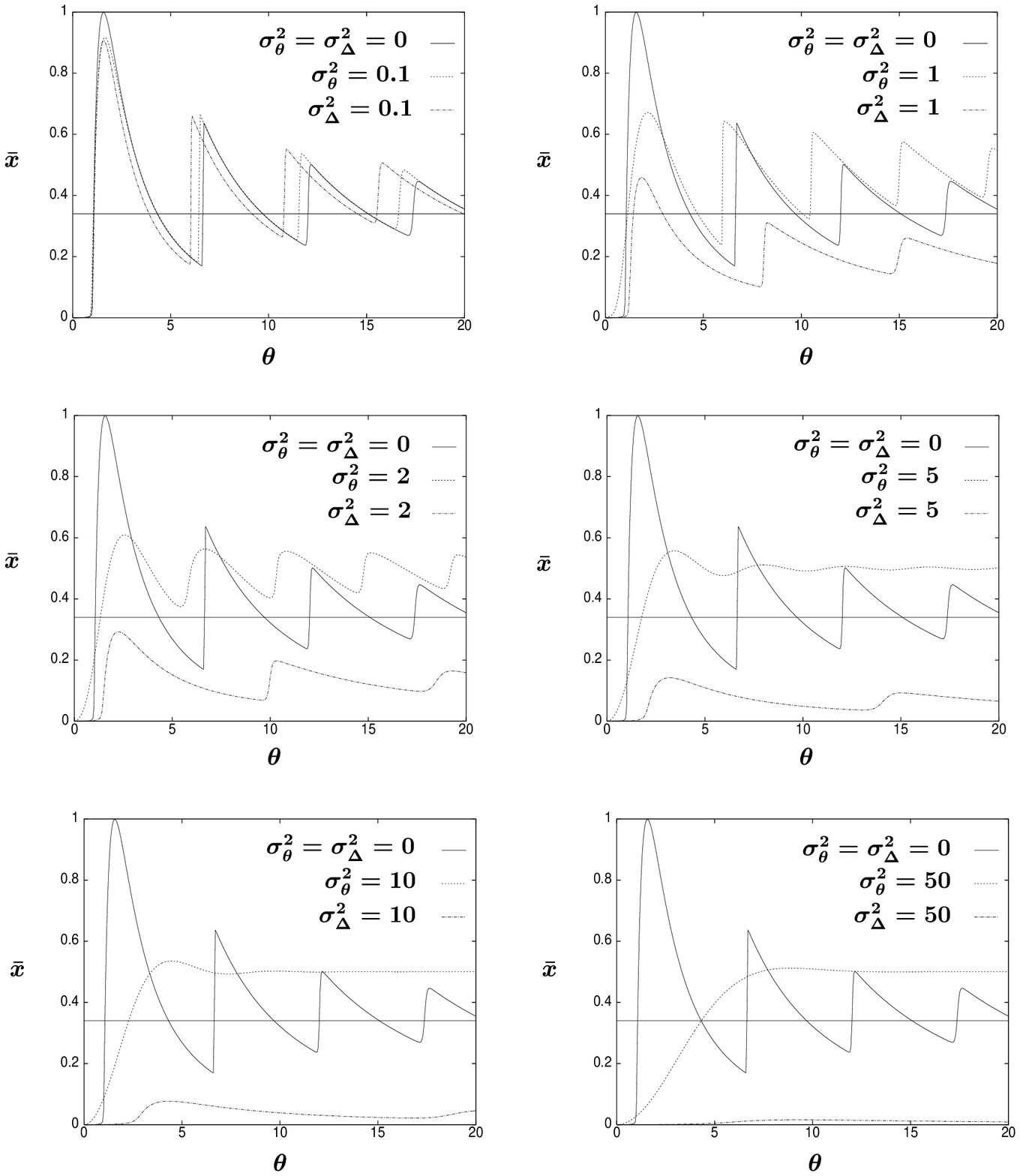}

\end{picture}  
\vspace{10cm} 
\figcap{ The normalized steady-state mean photon number ${\bar x}= {\bar n}/N$  
         as a function of $\theta$ for different values of
         $\sigma_{\theta}^2$ and $\sigma_{\Delta}^2$  
         when $a=1$, $n_b=0.15$, $\Delta=0$ and $N=1000$.   
         In the large $\theta$ and $N$ limit the order parameter approaches a constant value.  
         This asymptotic value is determined by \eq{hurra_formel} when  
         $\sigma_{\theta}^2=\sigma_{\Delta}^2=0$.   
         The solid line at ${\bar x}_{\infty} \approx 0.34$ in the
         figure corresponds to this asymptotic   
         value of the order parameter.  
\label{x_FIG} }  
\end{figure}

     Let us now consider the case $\sigma_{\theta}^2 \ll 1$. If, in addition, \eq{c_theta} is   
     satisfied, the quantity $\langle q(x) \rangle_{\theta}$ is reduced to \eq{q_n}.  
     If on the other hand \eq{c_theta} is not satisfied, i.e. $\theta \ll \sigma_{\theta}^2$, then   
     the saddle-point equation \eq{sadel_lign} has only the trivial solution for any   
     probability distribution $P_{\theta}(\xi)$ as it should.    
     The fluctuations in the atom cavity time  
     have then no dramatic effect on the order parameter ${\bar x}$ (see e.g. \fig{x_FIG}),   
     the excitation probabilities ${\cal P}(+)$ and ${\cal P}(+,+)$ or the phase diagram.

\vspace{0.8cm}  
  
\bc{  
\subsection{Detuning Fluctuations}  
\label{det_SEC}  
}\ec  
  
\vspace{0.0cm}

     In this Section we consider the situation when there is a spread in the detuning parameter $\Delta$.  
     Suppose that the spread in the detuning $\Delta$ is expressed in terms of a  stochastic variable   
     $\xi$ as described by the probability distribution $P_{\Delta}(\xi)$ such that   
     $\langle \xi \rangle = \Delta$ and $\langle ( \xi - \Delta )^2 \rangle = \sigma_{\Delta}^2$.   
     The averaged value of $q(x)$ with respect to $P_{\Delta}(\xi)$ is then 

\be \label{q_D_def}  
  \langle q(x) \rangle_{\Delta} = \int_{-\infty}^{\infty} d \xi \;   
   P_{\Delta}(\xi) \, \frac{x}{x+\xi^2} \sin^2 \left ( \theta \sqrt{x+\xi^2} \right )  ~~.  
\ee

    \noi  
    Non-trivial saddle-points of $V_0(x)$ can then be found by solving the equation

\be 
   1 = (2a-1) \, J_1( \theta,x(\theta) ) ~~,  
\ee

   \noi  
   where

\be  \label{J_1_def}  
   J_1(\theta,x)  = \theta^2 \; \int_{-\infty}^{\infty} d \xi \;  
   P_{\Delta}(\xi) \, \frac{ \sin^2 ( \, \theta \sqrt{x+\xi^2} \; ) }  
                                     {( \, \theta \sqrt{x+\xi^2} \; )^2 } \leq \theta^2 ~~.  
\ee

     For small values of $x$ we follow the argument of Section \ref{pump_SEC} concerning  
     the critical thermal-maser phase transition line. The critical transition line is then

\be \label{a_t_gen_II}  
  a(\theta) = \frac{1}{2} + \frac{1}{2} \frac{1}{J_1(\theta)} \leq 1  ~~,  
\ee

      \noi  
      where $J_1(\theta,0) \equiv J_1(\theta)$.    
      \eq{a_t_gen_II} coincides with the radius of convergence of the thermal probability   
      distribution \eq{p_n_geo}.  
      In general we must use a combination of analytical and numerical methods, as in the case   
      of $\sigma_{\theta}=0$ \cite{Rekdal99}, in order to get a detailed picture of the phase diagram.

        If the probability $P_{\Delta}(\xi)$ is sufficiently peaked such that the   
        its $\xi$-variation is fast in comparison to   
        $\xi$-variation of $\sin^2 \left ( \theta \xi \right )$,  
        i.e. if $\sigma_{\Delta} \theta \ll \pi/2$,  
        then \eq{J_1_def} is reduced to $J_1(\theta) =   [  \, 3 \, \sin^2(\theta \Delta)/\Delta^2   
        + \theta^2 \, \cos( 2 \theta \Delta ) - 2\theta \sin( 2 \theta \Delta )/\Delta  \, ]/\Delta^2$.  
        The first thermal-maser critical line is then given by

\be  \label{a_kons_Delta}  
  a(\theta) = \frac{1}{2} +   
              \frac{1}{2} \frac{\Delta^2}{ \sin^2( \theta \Delta ) + \sigma_{\Delta}^2 \, g(\theta) } ~~,  
\ee

    \noi   
    where $g(\theta) \equiv  3 \, \sin^2(\theta \Delta)/\Delta^2 + \theta^2 \cos( 2 \theta \Delta ) -   
    2\theta \, \sin( 2 \theta \Delta )/\Delta$, which is consistent with $\cite{Rekdal99}$ for   
    $\sigma_{\Delta}^2=0$.

    If, on the other  hand, the quantity $P_{\Delta}(\xi)$ has a   
    $\xi$-variation which is slow in comparison to the $\xi$-variation of   
    $\sin^2 ( \theta \xi )$, i.e. if $\sigma_{\Delta} \theta \gg \pi/2$, then   
    $J_1(\theta)$ reduces to

\be  \label{J_1_approx_stor}  
    J_1(\theta) = 1/2 \Delta^2 ~~.  
\ee

   \noi   
   \eq{a_kons_Delta} is then reduced to

\be  
   a(\theta) = \frac{1}{2} + \Delta^2  ~~.  
\ee

     In order to get explicit analytic results we now choose the following exponential probability  
     distribution for the detuning parameter

\be  
  P_{\Delta}(\xi) = \frac{1}{\sqrt{2\pi\sigma_{\Delta}^2}} \,   
  \exp \left (- \frac{(\xi-\Delta)^2}{2 \sigma^2_{\Delta}} \right ) ~~,   
\ee

   \noi  
   such that  $\langle \xi \rangle = \Delta$ and $\langle ( \xi - \Delta )^2 \rangle = \sigma_{\Delta}^2$.   
   Other choices are possible, but are not, as long as the distribution is sufficiently peaked,  
   expected to change the overall qualitative picture.      
   The averaged value of $q(x)$ with respect to $P_{\Delta}(\xi)$ is given by

\be \label{q_hele}  
  \langle q(x) \rangle_{\Delta} = \langle q(x) \rangle_0 + \langle q(x) \rangle_{osc} ~~,  
\ee

  \noi  
  where

\be \label{q_0}  
  \langle q(x) \rangle_0 \equiv \frac{1}{2} \frac{x}{\sqrt{2 \pi \sigma_{\Delta}^2}} \,   
        \int_{-\infty}^{\infty}  d \xi ~ \frac{ e^{- \frac{(\xi-\Delta)^2}{2 \sigma^2_{\Delta}} } }  
        {x+\xi^2}  ~~,  
\ee

  \noi and

\be \label{q_osc}  
  \langle q(x) \rangle_{osc} \equiv - \frac{1}{2} \frac{x}{\sqrt{2 \pi \sigma_{\Delta}^2}} \,   
                       \int_{-\infty}^{\infty}    
                       d \xi ~ \frac{ e^{- \frac{(\xi-\Delta)^2}{2 \sigma^2_{\Delta}} } }{x+\xi^2} \,    
                       \cos \left (  2 \theta \sqrt{x+\xi^2} \, \right ) ~~.  
\ee

     \noi  
     We observe that

\be  \label{int_omskr}  
   \langle q(x) \rangle_0 = \sqrt{\frac{x}{2 \sigma_{\Delta}^2}} ~  
   e^{-\Delta^2/2 \sigma_{\Delta}^2 + x/2 \sigma_{\Delta}^2}    
   \int_{\sqrt{x/2 \sigma_{\Delta}^2}}^{\infty}  d \eta ~ \exp{ \left [ \, -\eta^2 + \frac{1}{\eta^2}   
   \frac{\Delta^2}{2\sigma_{\Delta}^2} \frac{x}{2 \sigma_{\Delta}^2} \, \right ] } ~~.  
\ee


    \noindent \eq{int_omskr} can be used to find an expansion in the parameter $\Delta^2/2 \sigma_{\Delta}^2$.  
    For our purposes we notice that if $\sigma_{\Delta}^2$ is sufficiently large, i.e.

\be \label{stor_sigma}  
  \sigma_{\Delta}^2 \gg 1 ~~,  
\ee

    \noi  
    then \eq{int_omskr} is reduced to  $\langle q(x) \rangle_0 \lesssim \frac{1}{2}\sqrt{\pi \,   
    x/2 \sigma_{\Delta}^2} + {\cal O}( \, x/2 \sigma_{\Delta}^2 \, )$.  
    Any solution of the corresponding saddle-point equation is much less then unity  
    since $\langle q(x) \rangle_{osc} \leq \langle q(x) \rangle_0$.  
    The maser-maser phase transitions will therefore be less significant.  
    The phase diagram does then essentially consist of critical thermal-maser line(s) only.  
    For sufficiently large values of $\theta$, i.e.

\be  \label{stor_theta_her}  
  \theta \gg \frac{\sqrt{2 \pi \sigma_{\Delta}^2}}{2a-1} ~~,  
\ee

     \noi  
     the oscillating term can be estimated by $|\langle q(x) \rangle_{osc}| \approx   
     \left [ \, \pi/8 \sigma_{\Delta}^2 \, \right ]^{1/4} \sqrt{(2a-1)/(8 \sigma_{\Delta}^2 \theta)}$  
     for the solution

\be \label{x_j_spes_igjen}  
  {\bar x} = (2a - 1)^2 \frac{\pi}{8 \sigma_{\Delta}^2} ~~,  
\ee

     \noi  
     of the saddle-point equation.  \eq{x_j_spes_igjen} is much less then unity due to the  
     assumption \eq{stor_sigma}. The term $\langle q(x) \rangle_{osc}$ is negligible  
     in comparison to $\langle q(x) \rangle_0$ when \eq{stor_theta_her} is satisfied.   
     With regard to the observable ${\cal P}(+)$, the requirement   
     $\theta \gtrsim \theta_{\nu=1} = 2 \pi N \sqrt{ {\bar x} + \Delta^2}$ is compatible with  
     \eq{stor_theta_her} provided that $N$ is sufficiently large, i.e.

\be  
\label{Nverylarge}  
   N \gg \frac{4}{\pi^3} \, \frac{2 \sigma_{\Delta}^2}{(2a-1)^2}  ~~,  
\ee

     \noi  
     for the solution \eq{x_j_spes_igjen} of the saddle-point equation.  
     The equilibrium probability distribution is given by \eq{p_j_gauss}   
     with $x_j={\bar x}$ as in \eq{x_j_spes_igjen} and 
     $\sigma_x^2 = (2a-1) [ \, a + n_b(2a-1) \, ]  \pi/(4 N \sigma_{\Delta}^2)$. The probability   
     distribution ${\bar p}(x)$ is therefore peaked around  ${\bar x}$ provided that   
     ${\bar x} \gg \sigma_x$, i.e.

\be \label{N_sig_bet}  
  N \gg \frac{16 [ \, a + n_b(2a-1) \, ]}{(2a-1)^3 \pi} \; \sigma_{\Delta}^2 ~~.  
\ee

   \noi  
   The $\nu$:th revival of ${\cal P}(+)$ occurs in the region where $\theta$ is close to

\be  
  \theta_{\nu} = (2a-1) \pi \, \sqrt{ \frac{\pi}{2}} \, \frac{\nu N}{\sigma_{\Delta}} ~~,  
\ee

     \noi  
     according to \eq{overl_theta_nu}.  
     \eq{ikke_overlapp_bet_NY} implies that two consecutive revivals are separated provided that

\be  
\label{nulimit}   \nu \leq \sqrt{ \frac{N}{\sigma_{\Delta}^2} \; \frac{(2a-1)^3 \pi}{16 \, [a + n_b(2a-1)]} }   
            - \frac{1}{2} ~~,  
\ee

\begin{figure}[t]  
\unitlength=1mm  
\begin{picture}(100,80)(0,0)  
  
\includegraphics{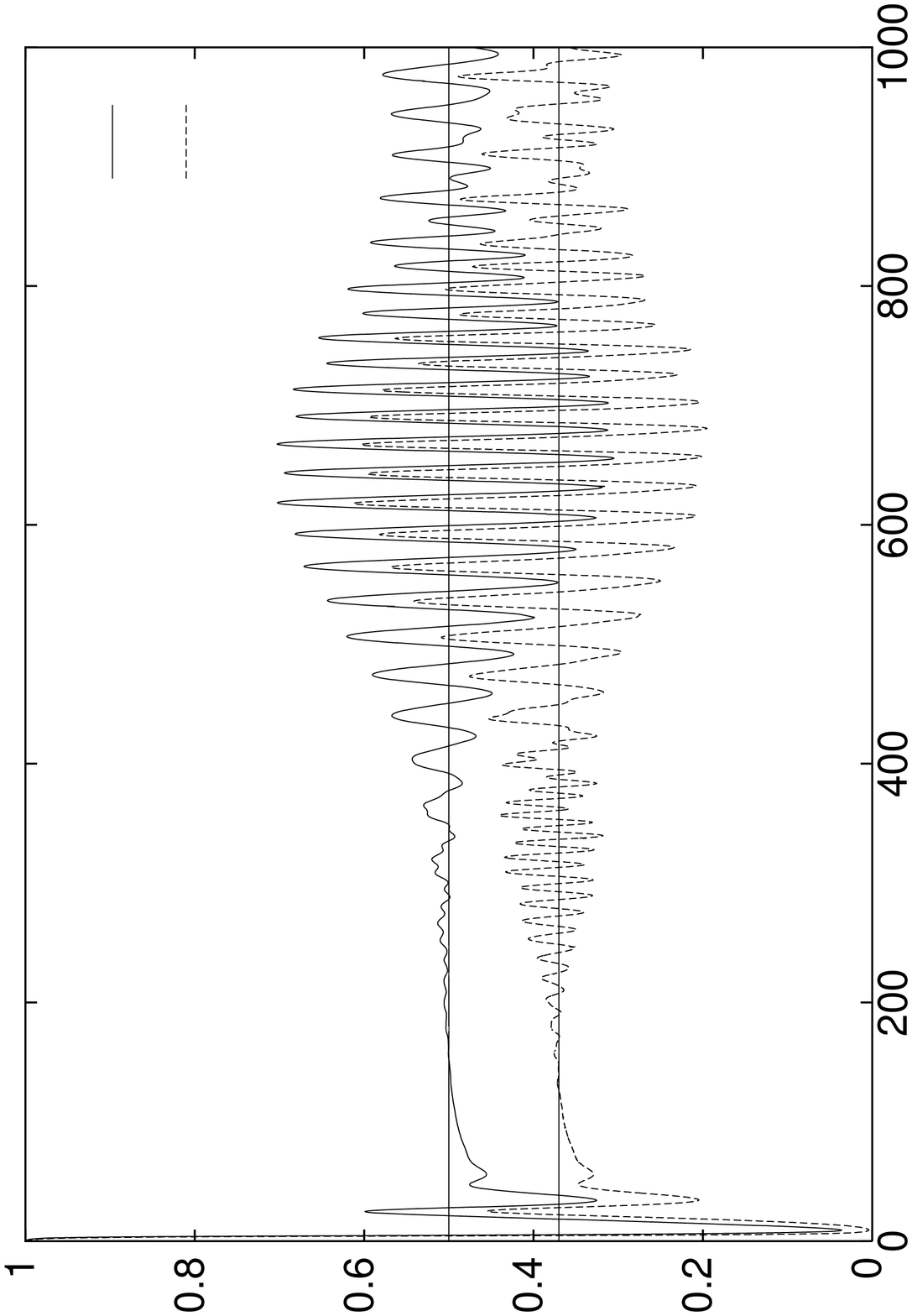}        
  
   \put(79,-6){\normalsize  \boldmath$\theta$}  
   
   \put(119.5,90.7){\normalsize  \boldmath${\cal P}(+)$}  
   \put(114,82){\normalsize  \boldmath${\cal P}(+,+)$}

\end{picture}  
\vspace{8mm}  
\figcap{ The probabilities ${\cal P}(+)$ (solid line) and ${\cal P}(+,+)$ (dotted line) as a   
         function of $\theta$ when the $q_m$-terms in \eq{p_n_eksakt} are averaged with respect to  
         $\Delta$ with $\sigma^2_{\Delta}=25$ and $N=800$.   
         All other parameters are as in \fig{p_pl_sig0_THETA_FIG}.   
         In the large $\theta$ and $N$ limit the probability ${\cal P}(+)$ approaches   
         ${\cal P}_{\infty}(+) = 0.5$ and the probability ${\cal P}(+,+)$ approaches   
         ${\cal P}_{\infty}(+,+) \approx 0.38$,  
         as indicated by a solid line in the figure. For sufficiently large $\theta$, ${\bar x} \approx 0.02$  
         and $\langle {\cal P}(+) \rangle \approx \langle {\cal P}(+,+) \rangle \approx 1$.  
\label{p_pl_sig25_DELTA_FIG} }  
\end{figure}

    \noi  
    which gives a bound on how many revivals that can be resolved in the excitation probability  
    ${\cal P}(+)$.   
    Fluctuations in the atom cavity transit time with a large $\sigma_{\theta}^2$ give, however,   
    an order parameter of the order one. This difference between $\theta$- and $\Delta$-fluctuations   
    is illustrated in \fig{x_FIG}. In \fig{p_pl_sig25_DELTA_FIG} we consider physical parameters such   
    that Eqs. (\ref{stor_sigma}), (\ref{Nverylarge}) and (\ref{N_sig_bet}) are satisfied.   
    We observe that ${\cal P}(+)$ and ${\cal P}(+,+)$ now exhibit  
    well pronounced revivals. With the same set of physical parameters one can also verify that the  
    approximative but analytical expression for ${\cal P}(+)$ as obtained by a Poisson resummation   
    technique, i.e. \eq{p_P_Poisson}, is in good agreement with the exact expression for   
    ${\cal P}(+)$. We also notice that according to \eq{nulimit}, only the first revival in   
    ${\cal P}(+)$ is clearly resolved in this case, which agrees well with our numerical results.

\begin{figure}[t]  
\unitlength=1mm  
\begin{picture}(100,80)(0,0)  
  
\includegraphics{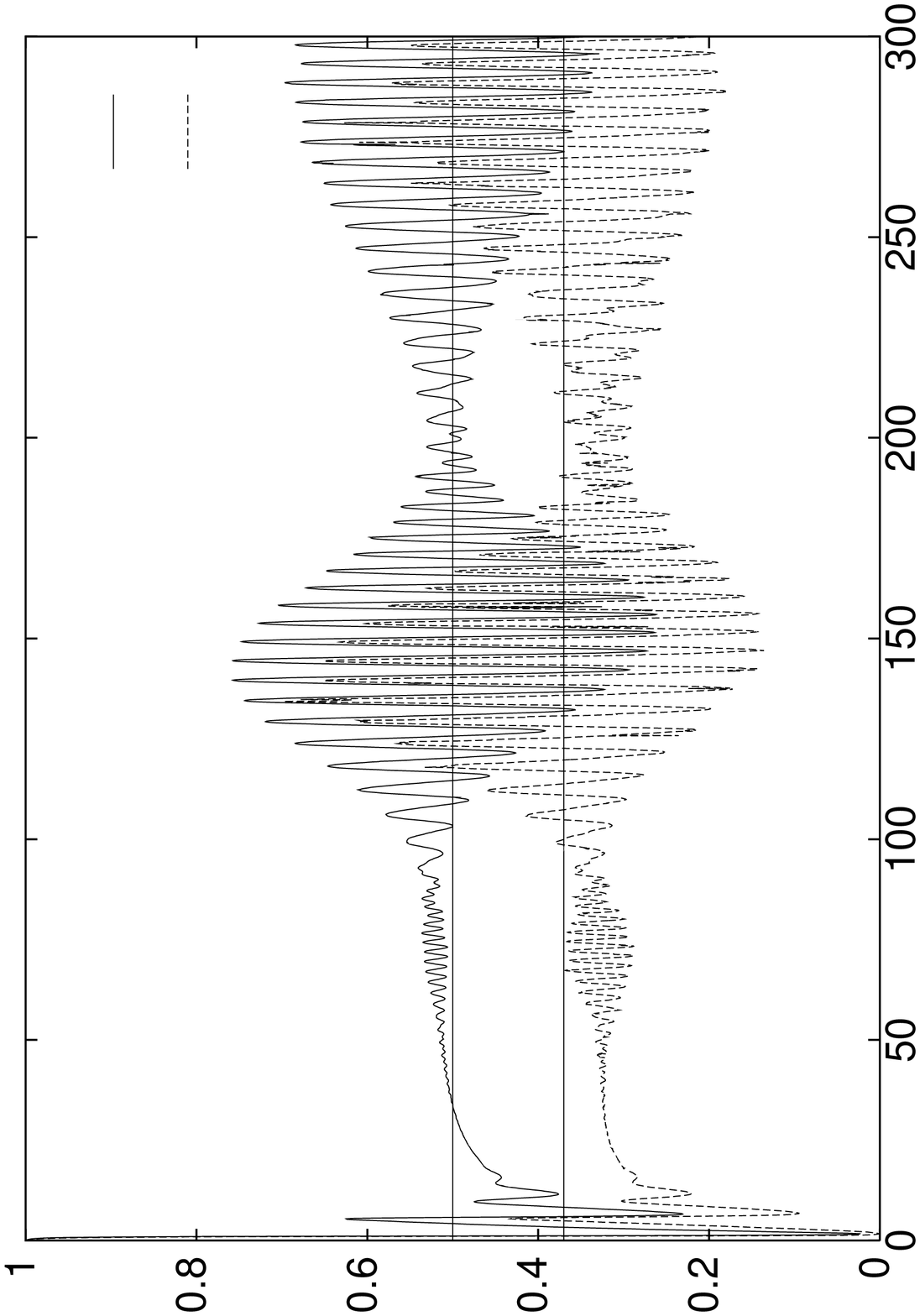}        
  
   \put(79.5,-6){\normalsize  \boldmath$\theta$}  
   
   \put(119.5,90.7){\normalsize  \boldmath${\cal P}(+)$}  
   \put(114,82){\normalsize  \boldmath${\cal P}(+,+)$}

\end{picture}  
\vspace{8mm}  
\figcap{ The probabilities ${\cal P}(+)$ (solid line) and ${\cal P}(+,+)$ (dotted line) as a   
         function of $\theta$ when the $q_m$-terms in \eq{p_n_eksakt} are averaged with respect to  
         $\Delta$ with $\sigma^2_{\Delta}=0.1$   
         All other parameters are as in \fig{p_pl_sig0_THETA_FIG}.  
         In the large $\theta$ and $N$ limit the probability ${\cal P}(+)$ approaches   
         ${\cal P}_{\infty}(+) = 0.5$ and the probability ${\cal P}(+,+)$ approaches   
         ${\cal P}_{\infty}(+,+) \approx 0.38$,  
         which is indicated by a solid line in the figure. For
         sufficiently large $\theta$, ${\bar x} \approx  
         0.5$, $\langle {\cal P}(+) \rangle \approx 0.5$ and 
         $\langle {\cal P}(+,+) \rangle \approx 0.38$.  
\label{p_pl_sig01_DELTA_FIG} }  
\end{figure}

    Let us now, on the other hand, consider the case when $\sigma_{\Delta}^2$ is small but non-zero, i.e.

\be \label{liten_sigma}  
  0 < \sigma_{\Delta}^2 \ll 1 ~~.  
\ee

     \noi  
     The quantity $\langle q(x) \rangle_0$ can then be approximated according to   
     $\langle q(x) \rangle_0 = x/[ \, 2(x + \Delta^2) \, ]$ and the term $\langle q(x) \rangle_{osc}$ has   
     an upper bound  $|\langle q(x) \rangle_{osc}| \leq 1/[ \, 1 + \epsilon^2 \, ]^{1/4}$, where  
     $\epsilon \equiv 2 \sigma_{\Delta}^2 \theta x/( \, x+\Delta^2 \, )^{3/2}$. For sufficiently  
     large values of $\theta$, i.e.

\be  \label{t_bet_zstor}  
   \theta \gg \frac{1}{2 \sigma_{\Delta}^2} ~ \frac{(a-1/2)^{3/2}}{a-1/2-\Delta^2} ~~,  
\ee

    \noi  
    the term $\langle q(x) \rangle_{osc}$ is negligible in comparison to $\langle q(x) \rangle_0$  
    for the solution

\be \label{x_j_nuh}  
  {\bar x} = a-1/2 - \Delta^2 ~~,  
\ee

     \noi  
     of the saddle-point equation.  
     The maser-maser phase transitions will therefore be less significant if Eqs. (\ref{liten_sigma}) and  
     (\ref{t_bet_zstor}) are satisfied.  
     The phase diagram does then essentially consist of critical thermal-maser line(s) only.  
     For sufficiently small values of $\theta$, i.e.

\be  
   \theta \ll \frac{1}{2\sigma_{\Delta}} ~~,  
\ee

     \noi  
     \eq{q_hele} is reduced to \eq{q_n}. Hence there is no dramatic change in the order  
     parameter or in the phase diagram. With regard to the observable ${\cal P}(+)$,   
     the requirement $\theta \gtrsim \theta_{\nu=1} = 2 \pi N \sqrt{{\bar x}+\Delta^2}$   
     is compatible with \eq{t_bet_zstor} provided that $N$ is sufficiently large, i.e.

\be  
\label{deltaNlarge}  
   N \gg \frac{1}{4 \pi \sigma_{\Delta}^2} \; \frac{\sqrt{a-1/2}}{a-1/2-\Delta^2} ~~.  
\ee

     \noi  
     The equilibrium probability distribution is then given by \eq{p_j_gauss}  
     with $x_j={\bar x}$ as in \eq{x_j_nuh} and $\sigma_x^2   
     = [ \, a + n_b(2a-1) \, ]/(2N)$ according to \eq{sigma_x}. The distribution ${\bar p}(x)$   
     is peaked around ${\bar x}$ provided that ${\bar x} \gg \sigma_x$, i.e.

\be \label{N_sig_bet_igjen}  
  N \gg \frac{1}{2} \; \frac{a + n_b(2a-1)}{a-1/2-\Delta^2} ~~.  
\ee

   \noi  
   The $\nu$:th revival of ${\cal P}(+)$ occurs in the region where $\theta$ is close to   
  
\be  
  \theta_{\nu} = 2 \pi \nu N \, \sqrt{a-1/2 } ~~,  
\ee

     \noi  
     according to \eq{overl_theta_nu}.  
     Furthermore, two consecutive revivals are separated provided that

\be  
   \nu \leq \sqrt{ \frac{N}{2} \; \frac{(2a-1)^2}{a + n_b(2a-1)} } - \frac{1}{2} ~~,  
\ee

     \noi  
     which gives a bound on how many revivals that can be resolved in the excitation   
     probability ${\cal P}(+)$.

\begin{figure}[t]  
\unitlength=1mm  
\begin{picture}(100,80)(0,0)

\includegraphics{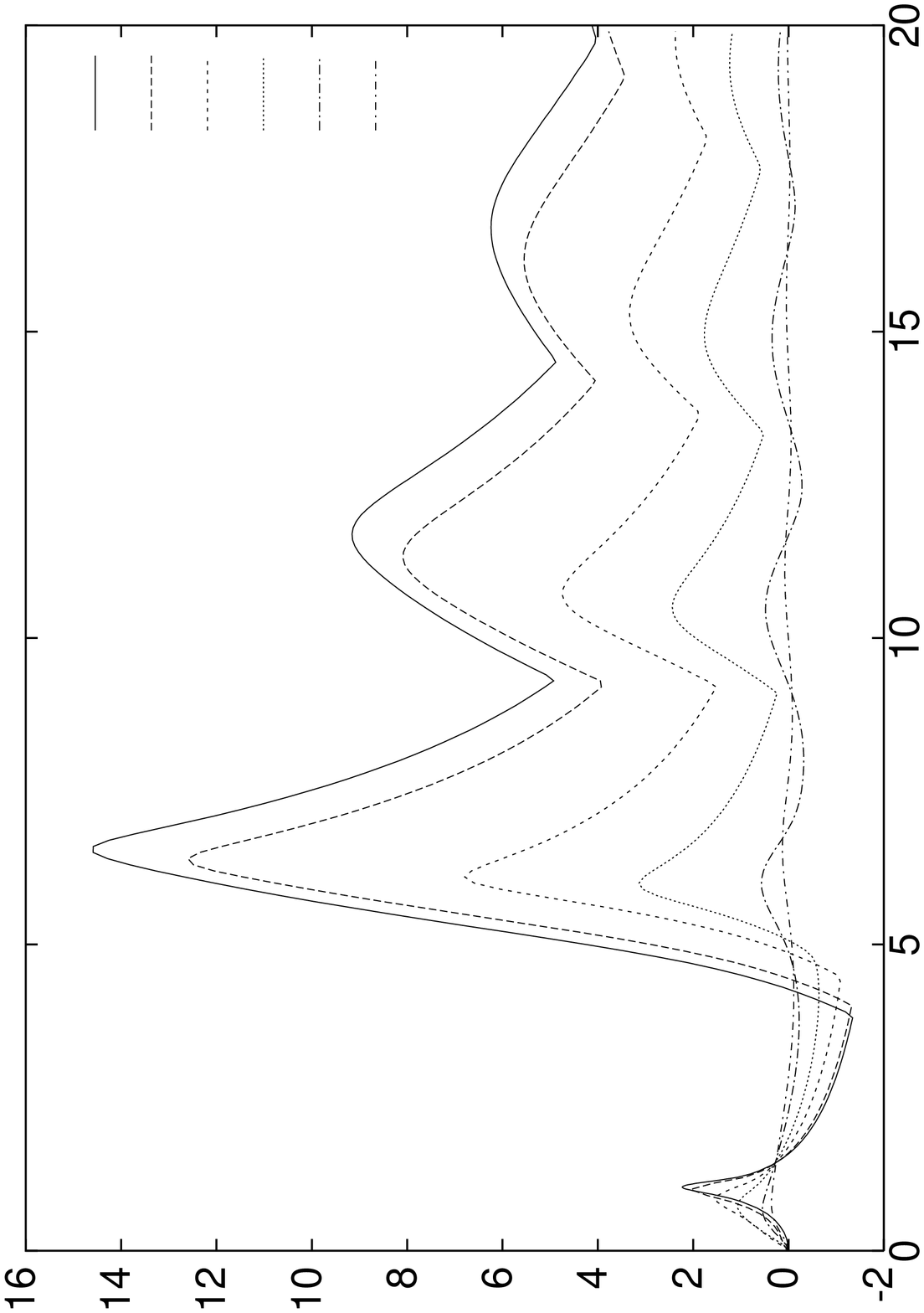}

   \put(119.5,76){\normalsize     \boldmath$\sigma_{\theta}^2=0$}  
   \put(116,69.5){\normalsize     \boldmath$\sigma_{\theta}^2=0.1$}  
   \put(116,63){\normalsize     \boldmath$\sigma_{\theta}^2=0.5$}  
   \put(119.5,57){\normalsize     \boldmath$\sigma_{\theta}^2=1$}  
   \put(116,50.5){\normalsize     \boldmath$\sigma_{\theta}^2=2.5$}  
   \put(119.5,44){\normalsize     \boldmath$\sigma_{\theta}^2=5$}

    \put(79.5,-24){\normalsize     \boldmath$\theta$}  
    \put(-13,36){\normalsize     \boldmath$\log( \gamma \xi )$}

\end{picture}  
\vspace{2.5cm}  
\figcap{ The logarithm of the correlation length  $\gamma \xi$ as a function of   
         $\theta$ for various values of $\sigma_{\theta}^2 = 0,0.1,0.5,1,2.5$ and $5$ when $N=100$,  
         $n_b=0.15$, $\Delta=0$ and $a=1$.   
\label{corr_FIG} }  
\end{figure}

       In \fig{p_pl_sig01_DELTA_FIG} we consider physical parameters such that Eqs. (\ref{liten_sigma}),   
       (\ref{deltaNlarge}) and (\ref{N_sig_bet_igjen}) are satisfied.  
       In this case we also observe well pronounced revivals in ${\cal P}(+)$ and ${\cal P}(+,+)$.   
       The width and location of the revival in ${\cal P}(+)$ agrees well with the approximative   
       expression \eq{p_P_Poisson}.

\vspace{0.5cm}  
\bc{  
\section{The Correlation Length}  
\label{corr_KAP}   
}\ec  
\vspace{0.5cm}

   Let us now consider long-time correlations in the large $N$ limit  
   as was first introduced in Ref.\cite{ElmforsLS95}.  
   These correlations are most  
   conveniently obtained by making use of the continuous time  
   formulation of the micromaser system \cite{Lugiato87}.   
   The lowest eigenvalue $\lambda_0=0$ of the matrix $L$, as defined in Section \ref{revivals_SEC},  
   then determines the stationary equilibrium solution vector $p={\bar p}$ as given by   
   Eq.~(\ref{p_n_eksakt}). The next non-zero eigenvalue $\lambda$ of $L$, which  
   we  ~determine ~numerically, ~will then  ~determine typical   
   scales for the approach to the stationary situation.  
   The joint probability for observing two atoms, with a time-delay $t$ between   
   them, can now be used in order to define a correlation length $\gamma^A(t)$ \cite{ElmforsLS95}.  
   At large times $t\rightarrow \infty$,   
   we define the atomic beam correlation length $\xi_A$ by \cite{ElmforsLS95}

\be  
   \gamma_A(t) \simeq e^{-t/\xi_A}~~,  
\ee

   \noi  
   which then  is determined by $\lambda$, i.e. $\gamma\xi_A=1/\lambda$.  
   For photons, we define a similar correlation length $\xi_{C}$. It  
   follows that the correlation lengths are identical, i.e.   
   $\xi_A=\xi_C\equiv \xi$ \cite{ElmforsLS95}.

   The correlation length $\gamma \xi$ is shown in Fig.~\ref{corr_FIG}  
   for different values of $\sigma_{\theta}^2$  
   and exhibits large peaks for the pump parameters   
   $\theta^*_{k k+1}$, $\theta_0^*$ and/or   
   $\theta^*_{tk}$ \cite{Rekdal99}.   
   The correlation grows at most as $\sqrt{N}$ at $\theta = \theta_0^*$.  
   At $\theta_{kk+1}^*$ and $\theta_{tk}^*$ the large $N$ dependence is, however, different.    
   At these values of the pump parameter   
   there is instead a competition between two neighbouring minima of $V_0(x)$.  
   Using the technique of \cite{ElmforsLS95} it can be  
   shown that the peak close to $\theta = \theta_{k k+1}^*$ is

\be   
  \gamma \xi \simeq e^{ N \Delta V_0} ~~,  
\ee

    \noi   
    where $\Delta V_0$ is the smallest potential barrier between the two competing minima   
    of the effective potential $V_0(x)$. For increasing values of $\sigma_{\theta}^2$ or $\sigma_{\Delta}^2$   
    this potential barrier $\Delta V_0$ decreases.  The corresponding correlation length will  
    then also decrease (see e.g. \fig{corr_FIG}).   
    Numerical study shows that adding fluctuations to the atom-photon frequency detuning  
    result in the same quantitative effect: the peaks in the correlation length decreases  
    due to smaller potential barriers $\Delta V_0$.

\vspace{1cm}  
\bc{  
\section{Conclusion}  
\label{final_KAP}   
}\ec  
\vspace{0.0cm}

    In conclusion, we have investigated various effects of including noise-producing mechanisms   
    in a micromaser system,   
    like a velocity spread in the atomic beam or a spread in the atom-photon frequency detuning.  
    For sufficiently large widths of the corresponding probability distributions, i.e.  
    for sufficiently large $\sigma_{\theta}^2$ or a non-zero $\sigma_{\Delta}^2$,   
    the excitation probabilities of atoms leaving the microcavity exhibit well-pronounced revivals.   
    Noise therefore tends to increase the order in the system.  
    Furthermore, the maser-maser phase transition disappear for sufficiently large   
    $\sigma_{\theta}^2$ or $\sigma_{\Delta}^2$,  
    in which case  the phase diagram in the $a$- and $\theta$- parameter space consists essentially   
    of a single maser-thermal critical line.   
    We have also shown that the correlation length drastically decreases when noise-producing   
    mechanisms are included in the system.

\newpage  
\vspace{1cm}  
\begin{center}  
{\bf APPENDIX}  
\end{center}  
\vspace{3mm}

     In this Appendix we derive an exact analytical expression  
     for $V_0(x)$ in the large $\theta$ limit when the quantity $q_n$ has no spread in any of the  
     physical parameters, i.e. when $q(x)$ is as given in \eq{q_n}.   
     The potential $V_0(x)$ may then be re-written in the from

\bea \label{mildert_denne}  
  V_0(x) &=& - \int_0^x  d \nu    
         \bigg \{ ~   
         \ln \left [ a \frac{\nu}{\nu + \Delta^2} + n_b \nu  \right ]  +    
         \ln \left [ 1 +\frac{a \sin^2( \theta \sqrt{\nu + \Delta^2})-a}{a + n_b( \nu + \Delta^2)} \right ]  
         \\ \nonumber &-&   
   \ln \left [ b \frac{\nu}{\nu + \Delta^2} + (1+n_b) \nu  \right ]  -  
   \ln \left [ 1 +\frac{b \, \sin^2( \theta \sqrt{\nu + \Delta^2})-b}{b + (1+n_b)( \nu + \Delta^2)} \right ]   
          \bigg \} ~~.  
\eea

   \noi  
   By making use of the power series expansion of the logarithm we obtain

\bea \label{V_expan}  
  V_0(x) &=& - \int_0^x  d \nu \, \bigg \{ ~  
         \ln \left [ \left( \; b \, \displaystyle{ \frac{\nu}{\nu + \Delta^2} } + (1+n_b) \nu \; \right )  
            /\left ( \; a \, \displaystyle{ \frac{\nu}{\nu + \Delta^2} } +    n_b  \nu \; \right )  \right ]    
         \\ \nonumber &+&  \sum_{n=1}^{\infty} \; \frac{1}{n}   
         \bigg [ ~ \left (\frac{a}{a+n_b(\nu+\Delta^2)} \right )^n  - 
                   \left (\frac{b}{b+(1+n_b)(\nu+\Delta^2)} \right )^n \; \bigg ]    
        \cos^{2n} ( \theta \sqrt{\nu + \Delta^2} ) ~ \bigg \} ~~.   
\eea

    \noi  
    In the large $\theta$ limit only the non-oscillating terms in \eq{V_expan} contribute, i.e.

\be \label{app_final}  
   V_0(x) = \int_0^x  d \nu \, \left \{ ~ \ln \left [ \frac{1-a}{a} \right ] +   
     f(\, y(\nu) \, ) - f( \, w(\nu) \,)   ~ \right \} ~~,  
\ee

    \noi   
    where   
  
\be  
  f(z) = \ln z + R(z) ~~,  
\ee

    \noi   
    and  $y(x) = a/( \, a+n_b(x+\Delta^2) \, )$ and $w(x) =   
    b/( \, b +(1+n_b)(x+\Delta^2) \, )$. The function $R(z)$ is defined by

\be \label{R_funksj}  
  R(z) = \sum_{n=1}^{\infty} \frac{1}{n} \frac{1}{2^{2n}}  \frac{(2n)!}{(n!)^2} \; z^n ~~,  
\ee

    \noi  
    which is a generalized hypergeometric series \cite{Magnus53}.  
   By differentiating \eq{app_final} we therefore, finally, obtain \eq{hurra_formel}.

\vspace{0.8cm}  
\newpage  
\begin{center}  
{\bf ACKNOWLEDGEMENT}  
\end{center}  
\vspace{0.3cm}

   The authors wish to thank the members of NORDITA  
   for the warm hospitality while the present work was completed.  
   The research has been supported in part  by the Research Council of  
   Norway under contract no. 118948/410.

\vspace{3mm}



\end{document}